\documentclass[twocolumn,prb,aps]{revtex4} 
\usepackage[dvips]{graphicx}
\usepackage{longtable}
\usepackage[fleqn]{amsmath}

\usepackage{color}
\begin{document}
\title{Complementary views on electron spectra:\\From Fluctuation Diagnostics to real space correlations}
\author{O.~Gunnarsson,$^1$ J.~Merino,$^2$ T.~Sch\"afer,$^{3,4,5}$ 
G.~Sangiovanni,$^6$ G.~Rohringer,$^7$ and A.~Toschi$^3$ } 
\affiliation{
$^1$ Max-Planck-Institut f\"ur Festk\"orperforschung, Heisenbergstrasse 1, D-70569 Stuttgart, Germany \\ 
$^2$ Departamento de F\'isica Te\'orica de la Materia Condensada, Condensed Matter\
Physics Center (IFIMAC) and Instituto Nicol\'as Cabrera, Universidad\
Aut\'onoma de Madrid, Madrid 28049, Spain\\
$^3$Institute of solid state physics, Vienna University of Technology, 1040 Vienna, Austria \\
$^4$ Centre de Physique Th{\'e}orique, {\'E}cole Polytechnique, CNRS,
Universit{\'e} Paris-Saclay, 91128 Palaiseau, France\\
$^5$Coll{\`e}ge de France, 11 place Marcelin Berthelot, 75005 Paris, France \\
$^6$ Institute for Theoretical Physics and Astrophysics, University of Würzburg, Am Hubland 97074 Würzburg, Germany \\
$^7$ Russian Quantum Center, Skolokovo (Moscow), Russian Federation}
\begin{abstract}

We study the relation between the microscopic properties of a many-body system 
and the electron spectra, experimentally accessible by photoemission. In a recent 
paper [Phys. Rev. Lett. {\bf 114}, 236402 (2015)], we introduced the ``fluctuation
diagnostics'' approach, to extract the dominant wave vector dependent bosonic 
fluctuations from the electronic self-energy. Here, we first reformulate the theory 
in terms of fermionic modes, to render its connection with resonance valence bond 
(RVB) fluctuations more transparent.  Secondly, by using a large-$U$ expansion, 
where $U$ is the Coulomb interaction, we relate the 
fluctuations to real space correlations.  Therefore, it becomes possible to study 
how electron spectra are related to charge, spin, superconductivity and RVB-like real 
space correlations, broadening the analysis of an earlier work [Phys.  Rev. B {\bf 89}, 
245130 (2014)]. This formalism is applied to the pseudogap physics of the two-dimensional 
Hubbard model, studied in the dynamical cluster approximation. We perform calculations 
for embedded clusters with up to 32 sites, having three inequivalent ${\bf K}$-points 
at the Fermi surface. We find that as $U$ is increased, correlation functions gradually
attain values consistent with an RVB state. This first happens for correlation functions 
involving the antinodal point and gradually spreads to the nodal point along the Fermi
surface. Simultaneously a pseudogap opens up along the Fermi surface. We relate this to 
a crossover from a Kondo-like state to an RVB-like localized cluster state and to the
presence of RVB and spin fluctuations. These changes are caused by a strong momentum
dependence in the cluster bath-couplings along the Fermi surface.  We also show, from a more 
algorithmic perspective, how the time-consuming calculations in fluctuation diagnostics 
can be drastically simplified. 
\end{abstract}
\maketitle
\section{Introduction}

The one-particle electronic spectrum is experimentally accessible, often with high accuracy, by means of 
photoemission and inverse photoemission spectroscopy.\cite{Hedin,Timusk1999,ARPESrev}
While the spectral functions contain important information about
the properties of the system, the extraction of this information is highly nontrivial. 
In fact, in the presence of strong correlations, different theoretical groups often obtain similar results
for spectra, but draw different conclusions about the underlying physics. There 
is therefore a great need for reasonably unbiased methods for determining
the physics from a spectral analysis.  
In particular, an important step would be to relate observable spectral properties 
to real space correlation functions and to momentum-dependent fluctuating modes.

Conceptionally, we can think of the electron spectrum $\rho({\bf k},\varepsilon)$
in two quite different ways. 
Within a standard quantum many-body formalism, we can write
\begin{eqnarray}\label{eq:1.1a}
&&\rho({\bf k},\varepsilon)=-{\frac{1}{\pi}} {\rm Im} g({\bf k},\varepsilon+i0^+) \\          
&&g({\bf k},z)={\frac{1}{z+\mu-\varepsilon_{\bf k}-\Sigma({\bf k},z)}}, \nonumber
\end{eqnarray}
where $g$ is the Green's function, $\Sigma$ is the self-energy, ${\bf k}$ is a wave vector, 
$\varepsilon$ the energy and $\mu$ is the chemical potential.  
This representation emphasizes the role of the self-energy and it allows, 
by applying the ``fluctuation diagnostics'' approach \cite{Fluct} to $\Sigma$, to identify the predominant
fluctuating modes of the system.

Alternatively, using a more quantum-chemistry oriented point of view, we can express the spectral function also as
\begin{eqnarray}\label{eq:1.2}
&&\rho({\bf k},\varepsilon)=\sum_n |\langle E_n(N-1)|c_{\bf k}|E_0(N)\rangle|^2  \\
&&\times \delta[\varepsilon-E_0(N)+E_n(N-1)+\mu]. \nonumber
\end{eqnarray}
This expression has for simplicity been written for zero temperature ($T=0$) 
and $\varepsilon\le 0$ (photoemission). Here $|E_0(N)\rangle$ is the initial
(ground-)state for $N$ electrons and $|E_n(N-1)\rangle$ is the $n$th excited 
final state for $N-1$ electrons, resulting in the photoemission process. 
$E_0(N)$ and $E_n(N-1)$ are the corresponding energies. This formalism tends 
to focus on the properties of the initial state, often 
described in terms of real space correlation functions, and how they influence the coupling to relevant final states. It
therefore connects the spectra to ground-state properties in a natural way.

These two approaches provide two complementary views on the electronic spectra. 
In this paper, we describe how they work for theoretical calculations of strongly-correlated systems and how they are interconnected.
In particular, we will show that $\rho$ can be related either to momentum-dependent fluctuations or to real space 
correlation functions. 

The spectra of strongly-correlated systems can be computed by means of different algorithms:
lattice QMC,\cite{bss} functional renormalization group,\cite{fRGrev} parquet approximation,\cite{parquet,parquet1,Yang2009}
and cluster extensions\cite{Maier} of the dynamical mean field theory (DMFT) \cite{DMFT, DMFTrev} 
such as the cellular-DMFT\cite{CDMFTlichtenstein,CDMFT} or the dynamical cluster 
approximation (DCA)\cite{DCA} as well as diagrammatic extensions 
\cite{DGA,DGA1,DF,DF1,nanoDGA,nanoDGA1,multiscale,1PI,DMF2RG,trilex,quadrilex} of DMFT. 

The relation between the complementary views given by Eq. (\ref{eq:1.1a}) and (\ref{eq:1.2}) will be illustrated 
using the example of the pseudogap\cite{Timusk} in the two-dimensional Hubbard model.
Because of its relevance for the physics of high-$T_c$ cuprates this problem has been extensively studied 
theoretically.\cite{Woelfle,Civelli,Macridin,Kyung,Ferrero,Imada,Millis,Liebscha,Sordi,Emery,Xu,Kosaka,Wang,Civelli2009,Yang,Gull}
In particular, the formation of the pseudogap has been ascribed to either spin \cite{Woelfle,Macridin,Kyung,Imada,Liebscha,Sordi}
or to superconducting fluctuations.\cite{Emery,Xu,Kosaka}
It has also been argued that the pseudogap and superconductivity phases can compete.\cite{Civelli2009,Yang,Gull}
The pseudogap is hence a perfect case of study where controversial interpretations of spectral properties can arise.

The interpretation of the pseudogap physics within the many-body framework of Eq.~(\ref{eq:1.1a}) 
has recently been discussed in Ref.~\onlinecite{Fluct} using the ``fluctuation diagnostics'' approach.
Within this scheme, the self-energy $\Sigma$ is first expressed, through the Schwinger-Dyson equation 
of motion (EOM), in terms of a two-particle scattering amplitude $F$. The EOM is then written in 
different representations, corresponding to the different scattering channels (charge, spin and particle-particle).
This way, it has been shown that spin fluctuations with ${\bf Q}=(\pi,\pi)$ gives the dominant 
($\sim$75$\%$) contribution to $\Sigma$, while pairing fluctuations do not contribute much for 
unconventional (non $s$-wave) superconductors. Such precise quantification has been possible by 
performing partial summations over the internal fermionic frequencies and momenta of the
EOM, and by comparing the different terms of the partial sum in the complementary representations.

While this approach provides us with a clear-cut ``diagnostics'' of the dominant fluctuations, it does
not explain \emph{why} they become large for given values of  the Coulomb interaction 
$U$ and certain momenta. Moreover, it does also not explain the reason for the strong momentum (${\bf K}$)
differentiation of $\Sigma$ occurring in the hole-doped region of the phase diagram of the two-dimensional Hubbard model. 

A rather different approach, in fact closer to the spirit of Eq.~(\ref{eq:1.2}),
was taken by Ferrero {\it et al.},\cite{Ferrero07} De Leo {\it et al.},\cite{Leo04} 
and Capone {\it et al.},\cite{Capone} who considered two- and three-orbital Anderson 
models and a four-impurity model.  Their work focuses on changes of the initial state, 
emphasizing the crossover from a Kondo-screened to an unscreened state. For the case 
of the A$_3$C$_{60}$ (A= K, Rb) compounds, they linked the appearance of a spectral 
pseudogap to an unscreened ground state.  In these works, the strong ${\bf K}$ 
dependence of the pseudogap, such as that of the cuprates, was however not addressed.

A similar formalism has been developed for analyzing the pseudogap in cuprates, 
and in particular its momentum  dependence.\cite{Jaime} For a small Coulomb interaction and for
an eight site embedded cluster in the DCA, it was found that cluster levels with the cluster 
momenta ${\bf K}=(\pi,0)$, $(0,\pi)$ and $(\pm \pi/2,\pm \pi/2)$ each forms a Kondo-like 
state with its bath. As $U$ is increased, the energy gain from the Kondo-like coupling is 
reduced and it becomes favorable to correlate (localize) electrons in the space of the $(\pi,0)$ 
and $(0,\pi)$ cluster levels, due to the energy gained when these levels form a localized state 
on the cluster.\cite{Jaime} At this point a pseudogap forms for ${\bf K}=(\pi,0)$ and 
$(0,\pi)$, due to the nondegenerate character of the localized state formed.\cite{Jaime} Due to the 
stronger coupling to the bath, the ${\bf K}=(\pm \pi/2,\pm\pi/2)$ levels at first stay 
mainly Kondo-like.\cite{Jaime} A further increase of $U$, however, eventually leads to 
a localization also of these levels, and the formation of a resonance valence bond\cite{Anderson} 
(RVB)-like state involving these levels and the $(\pi,0)$ and $(0,\pi)$ levels. This then 
leads to a pseudogap also  for ${\bf K}=(\pm \pi/2, \pm \pi/2)$.\cite{Jaime} While this approach 
gives insights into the structure of the ground-state correlations, 
in contrast to the fluctuation diagnostics, it does not provide a direct information about 
which fluctuations are important.

The purpose  of this paper is hence to merge the approach of Ref.~\onlinecite{Fluct} with that 
of Ref.~\onlinecite{Jaime}, and to clarify the intrinsic connection among the different results 
previously obtained by studying the physical mechanisms at work in the pseudogap phase.  In 
order to make the link between the two schemes, it is convenient to first reformulate the 
``fluctuation diagnostics'' of Ref.~\onlinecite{Fluct} in terms of fermionic modes. Technically, 
this corresponds to performing the partial summation in the EOM over the bosonic momenta and 
frequencies arriving at a decomposition of the self-energy in terms of fermionic contributions, 
rather than in terms of bosonic collective modes.  In a second step, we will make a large-$U$ 
approximation, which allows us to relate the bosonic- or fermionic-resolved fluctuations to real 
space correlations, e.g., charge, spin, superconductivity or RVB-like correlations.\cite{Anderson}

The work presented here has both physical and algorithmic implications.
In particular, the importance of RVB-like correlations found here as well as in Ref.~\onlinecite{Jaime}
is \emph{not} in contradiction to the importance of antiferromagnetic spin fluctuations found in
Ref.~\onlinecite{Fluct}. The formation of RVB-like correlations also leads to substantial
spin correlations,\cite{Jaime,JaimearX} which via the large $U$ approximation introduced
here, can be related to spin fluctuations.

On the numerical side, changing the order in performing the partial internal summations of the EOM 
represents a strong reduction of the numerical effort of both the ``original'' (bosonic) fluctuation 
diagnostics of Ref.~\onlinecite{Fluct} as well as of the  calculations presented in this paper, because 
some of the frequency sums can then be performed analytically.

In Sec.~\ref{sec:2} we present the methods and models. In Sec.~\ref{sec:3} we
develop the formalism. We give the relations between the vertex function and
the self-energy and show how the self-energy can be related to real space
correlations by making a large $U$ approximation. In Sec.~\ref{sec:4} correlation 
functions for the RVB-like state in isolated clusters are discussed. Finally, Sec.~\ref{sec:5} 
shows the relations between spectra and correlation functions. The results are 
summarized in Sec.~\ref{sec:6}. 

\section{Methods and models}\label{sec:2}

We use the dynamical cluster approximation (DCA)\cite{Maier} for calculating the self-energy, 
correlation functions and generalized susceptibilities. A cluster with $N_c$ sites is embedded 
in a self-consistent bath. We consider $N_c=4$, $8$ and $32$.  The cluster problem is solved 
using the Hirsch-Fye\cite{HirschFye} method. For the cases considered here there is no sign problem, 
which allows us to calculate the various quantities with a rather good statistical accuracy.

For most of the work here we use the two-dimensional Hubbard model.
This model is described by the Hamiltonian
\begin{eqnarray}\label{eq:2.11}
&&H =t \sum_{ \langle ij \rangle ,\sigma} (c^\dagger_{i \sigma}
c^{\phantom \dagger}_{j \sigma} + c^\dagger_{j \sigma} c^{\phantom \dagger}_{i \sigma})
-\mu\sum_{i\sigma}n_{i\sigma} + U \sum_{i} n_{i\uparrow} n_{i\downarrow}\nonumber \\
&&\equiv H_0+H_U, 
\label{eq:Hubbard}
\end{eqnarray}
where $\langle ij \rangle$ limits the sum to nearest neighbors, $\sigma$ is a spin 
index and $n_{i\sigma}=c^{\dagger}_ {i\sigma}c^{\phantom\dagger}_{i\sigma}$. The 
nearest neighbor hopping integral is $t$, the on-site Coulomb interaction is given by $U$ 
and we assume a square lattice. Typically a one-particle term 
\begin{equation}\label{eq:2.11a}
n_0U\sum_{i\sigma} n_{i\sigma} 
\end{equation}
is added to $H_0$ and subtracted from $H_U$, where $n_0$ 
is  an estimate of the occupancy per spin. This model is often used to describe cuprates 
superconductors.\cite{Dagotto,Scalapino} We use values of $U\sim (4-8)|t|$, which includes 
the range where a pseudogap occurs. The Coulomb interaction can also be expressed in 
reciprocal space
\begin{equation}\label{eq:2.12}
H_U={\frac{U}{N_c}}\sum_{{\bf K}_1{\bf K}_2 {\bf Q}}
c^{\dagger}_{{\bf K}_1\uparrow}c^{\phantom \dagger}_{{\bf K}_1+{\bf Q}\uparrow}
c^{\dagger}_{{\bf K}_2+{\bf Q}\downarrow}c^{\phantom \dagger}_{{\bf K}_2\downarrow}.
\end{equation}

Cuprates show a pseudogap around ${\bf K}=(\pi,0)$ and $(0,\pi)$ for moderate 
doping.  As the doping is reduced, the pseudogap extends over a larger range of 
${\bf K}$-values around the Fermi surface. A similar effect is seen at half-filling 
as $U$ is increased.\cite{Jaime,Schaefer2015} Here we therefore work at half-filling and 
vary $U$. This has the advantage that there is no sign problem, which makes it 
possible to study (DCA) clusters up to $N_c=32$. 

To improve our physical interpretation, below we also introduce a four-level 
model, which can be viewed as a simplified version of the $N_c=4$ case. The model is particularly 
transparent and it can be analyzed in great detail. We discuss the cluster in terms of the 
different ${\bf K}$-states. The Coulomb part can be written in the form
\begin{eqnarray}\label{eq:5.10a}
&&H_U=U_{xx}\sum_{\bf K} n_{{\bf K}\uparrow}n_{{\bf K}\downarrow}+U_{xy}
\sum_{{\bf K}\ne {\bf K}'} n_{{\bf K}\uparrow}n_{{\bf K}'\downarrow}   \\
&&+J\sum_{{\bf K}\ne {\bf K}'} c^{\dagger}_{{\bf K}\uparrow}c^{\dagger}_{{\bf K}'\downarrow}  
c^{\phantom \dagger}_{{\bf K} \downarrow}c^{\phantom \dagger}_{{\bf K}' \uparrow}
+J\sum_{{\bf K}\ne {\bf K}'} c^{\dagger}_{{\bf K}\uparrow}c^{\dagger}_{{\bf K}\downarrow} 
c^{\phantom \dagger}_{{\bf K}'\downarrow}c^{\phantom \dagger}_{{\bf K}' \uparrow}, \nonumber
\end{eqnarray}
where the first two terms describe the direct Coulomb interaction and
the following two are exchange terms. For the Hubbard model $U_{xx}=U_{xy}=J=U/N_c$, as given in Eq.~(\ref{eq:2.12}). 
For the $N_c=4$ case, the orbitals ${\bf K}=(\pi,0)$ and $(0,\pi)$ are particularly 
important for the physics, since at half-filling they are at the chemical potential. 
Therefore, in building up our simplified model, we consider a Hamiltonian where only these 
two orbitals are included, while the mainly occupied orbital ${\bf K}=(0,0)$ as well as the 
mainly unoccupied ${\bf K}=(\pi,\pi)$ orbital are neglected. The two orbitals ${\bf K}=(\pi,0)$ 
and $(0,\pi)$ are connected to one bath orbital each, giving a four-level model. The level 
structure of the isolated cluster containing just the $(\pi,0)$ and $(0,\pi)$ levels can be 
fitted to the level structure of four isolated sites, relevant for the $N_c=4$ calculation. 
We then find that $U_{xx}$ has to be chosen slightly smaller than $U_{xy}$.\cite{Jaime} 
To simulate the $N_c=4$ calculation we then choose
\begin{equation}\label{eq:1.1}
U_{xx}=U-\Delta U,\hskip0.2cm U_{xy}=U+\Delta U,\hskip0.2cm J=U.
\end{equation} 
Here we have redefined $U$ ($U/N_c \to U$) to simplify the notations. Eq.~(\ref{eq:1.1}) 
violates rotational invariance in spin-space and this is the price to be paid for 
simulating a four-site cluster with two levels.  We use a very small $\Delta U$ and 
therefore this violation is small. 

For the four-level model we introduce the following one-particle Hamiltonian
\begin{eqnarray}\label{eq:5.1}
H_0&&=(\varepsilon_c-\mu)\sum_{i=1}^2\sum_\sigma n_{ic\sigma}+(\varepsilon_b-\mu)\sum_{i=1}^2\sum_\sigma n_{ib\sigma}
 \nonumber \\
&&+\sum_{i \sigma}V  (c_{ic\sigma}^{\dagger}c_{ib\sigma}^{\phantom \dagger}+c_{ib\sigma}^{\dagger}c_{ic\sigma}^{\phantom \dagger}).
\end{eqnarray}
where $c$ refers to cluster levels and $b$ to bath levels. We use the Coulomb 
interaction in Eq.~(\ref{eq:5.10a}), where ${\bf K}$ now runs over two levels 
($i=1,2$), describing ${\bf K}=(\pi,0)$ and $(0,\pi)$ respectively.

Although the $N_c=4$ case is simulated by $\Delta U>0$, we also consider $\Delta U<0$. 
The change from $\Delta U>0$ to $\Delta U<0$ leads, in fact, to a dramatic change of the 
ground-state and of the spectral function. This illustrates nicely how the spectral 
function can provide important information about the ground-state.

As we already mentioned, the coupling of the embedded cluster to the bath plays an important 
role, and therefore we discuss this here in a more general fashion. The 
noninteracting Green's function can be written as
\begin{equation}\label{eq:2.4a}
g_0({\bf K},i\nu_n)\equiv [i\nu_n-\Delta-\bar \varepsilon_{\bf K}-\Gamma_{\bf K}(i\nu_n)]^{-1},
\end{equation}
where $\nu_n$ is a Matsubara frequency, $\Delta=n_0U-\mu$
and $n_0$ was defined below Eq.~(\ref{eq:2.11a}).
Here the ${\bf k}$-space has been divided in $N_c$ patches around 
each cluster ${\bf K}$-vector, where each patch contains $\tilde 
N$ ${\bf k}$ states. We define $\bar \varepsilon_{\bf K} ={\frac{1}{N}}\sum_{\tilde {\bf k}} \varepsilon_{\bf K+\tilde k}$. 
The hybridization function $\Gamma$ satisfies an important sum 
rule\cite{ErikDCA}
\begin{equation}\label{eq:2.6}
-{\frac{1}{\pi}}\int {\rm Im} \Gamma_{\bf K}(\varepsilon+i0^{+})d\varepsilon ={\frac{1}{\tilde N}} 
\sum_{\tilde {\bf k}}(\varepsilon_{\bf K+\tilde k}-\bar \varepsilon_{\bf K})^2,
\end{equation}
where $\Gamma_{\bf K}(i\nu_n)$ has been analytically continued to real $\varepsilon$ 
and $0^{+}$ is an infinitesimal positive number. This result shows that 
the second moment of the $\varepsilon_{\bf k}$ inside a patch is a measure of the coupling to 
the bath. Due to the weak dispersion around ${\bf K=}(\pi,0)$, 
$| {\rm Im} \Gamma_{\bf K}(i\nu_n)|$ is much smaller for $(\pi,0)$  than 
for $(\pi/2,\pi/2)$.  The corresponding values are reported in Table~\ref{table:6.5} for $N_c=8$ and 32. 
For $N_c=8$ the difference is about a factor of four and for $N_c=32$ 
the difference is even larger. In the latter case the patch around ${\bf K}=(\pi,0)$
is very small and dominated by the saddle point for this ${\bf K}$.
\begin{table}
\caption{\label{table:6.5}Sum rule $-(1/\pi)\int {\rm Im}\Gamma({\bf K},\varepsilon+i0^{+}) d\varepsilon$
according to Eq.~(\ref{eq:2.6}) for different values of ${\bf K}$ and  $N_c$ and
for  $t=-0.25$.
}
\begin{tabular}{rrr}
\hline
\hline
\multicolumn{1}{c}{${\bf K}$} & \multicolumn{2}{c}{Sum rule }\\
  &  $N_c=8$ & $N_c=32$ \\
\hline
$(\pi,0)$ & 0.033 &0.002 \\
$(3\pi/4,\pi/4)$ & & 0.025 \\
$(\pi/2,\pi/2)$ & 0.149 & 0.047 \\
\hline
\hline
\end{tabular}
\end{table}

\section{Formalism and derivations}\label{sec:3}

\subsection{Two-particle vertex function}\label{sec:3.1}

We first introduce generalized susceptibilities in the Matsubara formalism 
for a finite temperature $T=1/\beta$. We follow the notations of Rohringer 
{\it et. al}.\cite{Rohringer}.
\begin{eqnarray}\label{eq:2.1}
&&\chi_{\sigma \sigma'}(k;k';q)          
=\int_0^{\beta}d\tau_1 \int_0^{\beta}d\tau_2 \int_0^{\beta}d\tau_3  \nonumber  \\
&& \times e^{-i[\nu\tau_1-(\omega+\nu)\tau_2+(\omega+\nu')\tau_3]}    \\
&& \times \langle T_{\tau}[c^{\dagger}_{{\bf k}\sigma}(\tau_1) c_{{\bf
    k}+{\bf q}\sigma}(\tau_2) c^{\dagger}_{{\bf k}'+{\bf
    q}\sigma'}(\tau_3) c_{{\bf k}'\sigma'}]\rangle  \nonumber \\ 
&& - \beta \, g_{\sigma}(k)g_{\sigma'}(k') \, \delta_{q=0}.\nonumber
\end{eqnarray}
Here we use the condensed notations $q=({\bf Q},\omega)$ and $k=({\bf
  K},\nu)$, where ${\bf Q}$ and ${\bf K}$ are (cluster) wave vectors and
$\omega$ and $\nu$ are Matsubara boson and fermion frequencies, respectively.
The Green's function is given by
\begin{equation}\label{eq:2.2}
g_{\sigma}(k)=-\int_0^{\beta} d\tau e^{i\nu \tau}\langle c_{{\bf K}\sigma}^{\phantom \dagger}(\tau)
c_{{\bf K}\sigma}^{\dagger} \rangle,
\end{equation}
where $c_{{\bf K}\sigma}^{\dagger}$ creates an electron with the wave vector
${\bf K}$ and spin $\sigma$ and $\langle .. \rangle$ is the thermodynamical average.
From $\chi$ we obtain the full vertex $F$
\begin{eqnarray}\label{eq:2.3}
&&\chi_{\sigma\sigma'}(k;k';q)=-\beta g_{\sigma}(k)g_\sigma(k+q)\delta_{kk'}\delta_{\sigma\sigma'} \\
&&-g_{\sigma}(k) g_{\sigma}(k+q)F_{\sigma\sigma'}(k;k';q)g_{\sigma'}(k') g_{\sigma'}(k'+q). \nonumber
\end{eqnarray}
In the SU(2) symmetric cases considered here, $g$ is independent of
$\sigma$ ($g=g_\uparrow=g_\downarrow$), and the index $\sigma$
is therefore dropped in the following.
The self-energy $\Sigma$ can be obtained from the equation of motion.
For the Hubbard model [Eqs.~(\ref{eq:2.11}, \ref{eq:2.12})] we have
\begin{equation}\label{eq:2.4}
\Sigma(k) 
=[g(k)]^{-1}\int_0^{\beta} \langle [H_U(\tau),c_{{\bf K}\sigma}(\tau)]c_{{\bf K}\sigma}^{\dagger}\rangle d\tau e^{i\nu \tau}.
\end{equation}
The commutator can be expressed in terms of the vertex function $F$,
leading to the Schwinger-Dyson equation        
\begin{eqnarray}\label{eq:2.9}
&&\Sigma(k)-({n\over 2}-n_0)U            \\
&&=-{U\over \beta^2N_c}\sum_{k',q}F_{\uparrow \downarrow}(k,k',q)    
g(k')g(k'+q)g(k+q)  \nonumber
\end{eqnarray}
where $N_c$ is the number of  ${\bf K}$-points in the embedded cluster. 

Exploiting the SU(2) symmetry of the Hubbard model for the paramagnetic state\cite{Rohringer} 
and ``crossing relations'',\cite{Rohringer} due to the electrons being identical 
particles, specific identities between different vertex functions can be
derived. By means of  these relations Eq.~(\ref{eq:2.9}) can be rewritten as\cite{Fluct}
\begin{eqnarray}\label{eq:2.10} 
&&\Sigma(k) - ({n\over 2}-n_0)U \\ 
&&=\hskip-0.0cm    
 {U\over \beta^2 N_c}\sum_{k',q} \, F_{sp}(k,k'; q) \, g(k')g(k'\! +\!q)g(k \!+ \!q), \nonumber  \\
      && =\hskip-0.0cm 
 -{U\over \beta^2 N_c}\sum_{k',q} \, F_{ch}(k,k'; q) \,
 g(k')g(k'+q)g(k+q), \nonumber \\
          && =\hskip-0.0cm
  -{U\over \beta^2 N_c}\sum_{k',q} \, F_{pp}(k,k';q) \, g(k')g(q-k')g(q-k), \nonumber
\end{eqnarray} 
where
\begin{eqnarray}\label{eq:2.10a}
&&F_{\rm ch}(k;k';q)=F_{\uparrow \uparrow}(k;k';q)+F_{\uparrow \downarrow}(k;k';q) \nonumber \\
&&F_{\rm sp}(k;k';q)=F_{\uparrow \uparrow}(k;k';q)-F_{\uparrow \downarrow}(k;k';q)  \\
&&F_{\rm pp}(k;k';q)=F_{\uparrow \downarrow}(k;k';q-k-k'). \nonumber 
\end{eqnarray}
The three equations in Eq.~(\ref{eq:2.10}) are all exact and give identical results 
if the summations are performed until convergence. Using the scheme
coined ``fluctuation diagnostics'',\cite{Fluct} 
 the partial contribution to these sums were studied as a
 function of $q$.\cite{Fluct,Kozik_Fluct} 
 In particular,  we recall that if there are low-lying spin fluctuations for ${\bf Q}={\bf Q}_0$, these give large contributions in the top 
formula in Eq.~(\ref{eq:2.10}) for ${\bf Q}={\bf Q}_0$ and small $\omega$. In the other
two formulas, on the other hand, the contributions are spread out over many ${\bf Q}$ 
and $\omega$. In a similar way, one can detect the presence of well defined charge and superconductivity fluctuations
from the second and third formula, respectively. This makes it possible to identify 
which fluctuations are important in determining a given numerical
result for the self-energy.  From the
algorithmic point of view, it is important to recall that the
advantage of this procedure w.r.t. a more direct decomposition\cite{Parquetprb} of the self-energy in terms of the parquet
equations,\cite{Bickersbook} is to avoid, at any step of the algorithm, the calculations
of possibly
divergent\cite{Schaefer2013,Janis2014,Kozik2015,Stan2015,Ribic2016,Schaefer2016,Tarantino2017,Gunnarsson2017,Vucicevic2017,Chalupa2017}
  two-particle irreducible vertex functions. 

Complementarily to the fluctuation diagnostics, in Ref.~\onlinecite{Jaime} the correlation function 
\begin{equation}\label{eq:3.1a}
 L( {\bf K })=\langle n_{{\bf K} \uparrow} n_{{\bf K} \downarrow} \rangle 
- \langle n_{{\bf K}\uparrow}  \rangle \langle n_{{\bf K} \downarrow} \rangle.
\end{equation}
was introduced to relate the spectral function to the underlying ground-state properties. 
This correlation function describes the transition from a Kondo-like state to localized 
state on the cluster.  For small $U$, the Kondo screening of the different ${\bf K}$ states 
 is important, as discussed in the introduction. We then obtained $L<0$, showing the 
 beginning of the formation of a spin $1/2$ state in the orbital ${\bf K}$. This spin 
 state was  found to couple antiferromagnetically to the bath, leading to a Kondo-like 
 state. For larger values of $U$, it was found that localized states form on the cluster, 
 e.g., for $N_c=4$ as in Eq.~(\ref{eq:5.2}) below. This leads to $L>0$.
 We will use this correlation function later,  
to clarify the relation between the results obtained by means of the
fluctuation diagnostics and the complementary approach of 
Ref.~\onlinecite{Jaime}.

\subsection{Relation to susceptibilities}\label{sec:3.2}

In this subsection,  we  start by reformulating  the  fluctuation
diagnostic approach of Ref.~\onlinecite{Fluct} in terms of fermionic
modes. This reformulation is, from a physically point of view,
rigorously equivalent to the bosonic one of Ref.~\onlinecite{Fluct}
for the case of SU(2)-symmetric models mostly considered in this work.
It allows however, to establish in a more immediate way the connection
 between the predominant fluctuations and the underlying correlations
 in real space. 

To this aim, we will introduce an extended set of correlation functions
designed to capture the complementary aspects of the underlying physics.
In particular, going beyond the derivations of  Ref.~\onlinecite{Fluct}, we will also study 
the ${\bf K}'$ dependence of the two-particle correlation function
after a summation over the transfer momentum ${\bf Q}$ has been 
performed, as there the connection to RVB-like correlations is more easily visible. 
At the same time, we will discuss how the extended formalism allows
for a drastic numerical simplification of the formulas previously used in 
fluctuation diagnostics.\cite{Fluct}

The fermionic reformulation of the equation of
Ref.~\onlinecite{Fluct}  is readily  obtained by following an alternative route in treating the equation of 
motion. This corresponds, in practice, to
the situation in which the frequency summations in 
Eqs.~(\ref{eq:2.9}, \ref{eq:2.10}) have already been performed. 
From Eq.~(\ref{eq:2.3}), 
we can see that the  susceptibility is the vertex function $F_{\uparrow\downarrow}$ 
times four Green's functions, while the contribution to the self-energy is 
$F_{\uparrow\downarrow}$ times three Green's functions. The derivation below essentially replaces 
the vertex function by the susceptibility divided by a Green's function and 
with two frequency summations performed. 

As for the derivation, we start by inserting the expression of the
Hubbard interaction in the commutator in Eq.~(\ref{eq:2.4}),
obtaining 
\begin{eqnarray}\label{eq:11}
&&\Sigma(k)+n_0U
=-{U \over N_cg(k)}\sum_{{\bf K}'{\bf Q}} \\
&& \times \int_0^{\beta}d\tau e^{i\nu \tau}\langle c_{{\bf K}+
{\bf Q}\uparrow}(\tau)c^{\dagger}_{{\bf K}'+{\bf Q}\downarrow}(\tau)
c^{\phantom \dagger}_{{\bf K}'\downarrow}(\tau)
c_{{\bf K}\uparrow}^{\dagger}\rangle . \nonumber
\end{eqnarray}
We now introduce a specific two-particle correlation function 
\begin{eqnarray}\label{eq:11a}
&&A_{\sigma \sigma'}({\bf K},{\bf K}',{\bf Q}) \\
&&=-{U\over N_c}\int_0^{\beta}d\tau e^{i\nu \tau}\langle c_{{\bf K}+
{\bf Q}\sigma}(\tau)c^{\dagger}_{{\bf K}'+{\bf Q}\sigma'   }(\tau)
c^{\phantom \dagger}_{{\bf K}'\sigma'   }(\tau)
c_{{\bf K}\sigma}^{\dagger}\rangle\nonumber  \\
&&={U\over N_c}\int_0^{\beta}d\tau e^{-i\nu \tau}
\langle c_{{\bf K}\sigma}^{\dagger}(\tau)c^{\phantom \dagger}_{{\bf K}+{\bf Q}\sigma}  
c^{\dagger}_{{\bf K}'+{\bf Q}\sigma'}c^{\phantom \dagger}_{{\bf K}'\sigma'} \rangle\nonumber,
\end{eqnarray}
and express Eq.~(\ref{eq:11}) in terms of $A_{\uparrow\downarrow}({\bf K},{\bf K}',{\bf Q})$.
In the following we focus on the lowest Matsubara fermion frequency
($\nu=\pi/\beta$) only;  this frequency will therefore not be shown
explicitly in the corresponding notation for $A_{\uparrow\downarrow}({\bf K},{\bf K}',{\bf Q})$. 
Comparing with Eqs.~(\ref{eq:2.3}, \ref{eq:2.9}), we can see that the integral in Eq.~(\ref{eq:11},
\ref{eq:11a}) corresponds, to a large extent, to the susceptibility $\chi_{\uparrow \downarrow}(k,k',q)$ summed over the 
frequencies $\nu'$ and $\omega$:

\begin{eqnarray}
&&\Sigma(k)+n_{0}U=   \\
&&\frac{1}{g(k)}\sum\limits_{{\bf K'}{\bf Q}}
  \frac{U}{\beta^{2}N_{c}}\sum\limits_{\nu'\omega}\left[\chi_{\uparrow\downarrow}(k,k',q)+\frac{N_{c}n}{2}g(k)\delta_{{\bf Q}{\bf 0}}\right] \nonumber \\
&&=\frac{1}{g(k)}\sum\limits_{{\bf K'}{\bf Q}}A_{\uparrow\downarrow}({\bf K},{\bf K'},{\bf Q})\label{eq:dse_a} \nonumber
\end{eqnarray}
Since all frequency summations have already been performed in Eq.~(\ref{eq:dse_a}), this expression 
is much easier to calculate than the corresponding
terms appearing in Eq.~(\ref{eq:2.10}). 
It is important to stress that this represents a {\sl significant  advantage} for the algorithms
based on partial summation of the EOM, making it possible, e.g., to perform
the fluctuation diagnostics of the DCA self-energy for much larger clusters ($N_c=32$) than before ($N_c=8$). 

Note that in Eq.~(\ref{eq:2.3}) a Hartree term was separated out, while it is kept in the definition
of $A_{\uparrow\downarrow}$ in Eq.~(\ref{eq:11a}), making $A_{\uparrow\downarrow}/g$ slightly 
different from $F_{\uparrow\downarrow}ggg$.  Similarly, $A_{\uparrow \uparrow}/g$ 
differs by a constant term from $F_{\uparrow\uparrow}ggg$.  However,
in this work we consider a half-filled  system and ${\bf K}$ at the
Fermi surface. For this case $g(k)$ is imaginary, and the difference in 
definition shows up in the imaginary part only. Hence, since we are mostly interested in the imaginary part of 
$\Sigma$ at the Fermi level, we can  focus on Re $A_{\sigma \sigma'}$
only, for which the difference between $A_{\uparrow\downarrow}/g$
and $F_{\uparrow\downarrow}ggg$ is of no particular interest. 

In order to rationalize the different, complementary treatments of the
EOM discussed in this work, we introduce the following correlation
functions:
\begin{eqnarray}\label{eq:11c}
&&B_{\sigma \sigma'}({\bf K},{\bf K}')=\sum_{\bf Q}A_{\sigma \sigma'}({\bf K},{\bf K}',{\bf Q}) \nonumber \\
&&C_{\sigma \sigma'}({\bf K},{\bf Q})=\sum_{{\bf K}'}A_{\sigma \sigma'}({\bf K},{\bf K}',{\bf Q}) \\
&&D_{\sigma \sigma'}(K)+n_0Ug(k) \nonumber \\
&&=\sum_{{\bf K}'}B_{\sigma \sigma'}({\bf K},{\bf K}')=\sum_{\bf Q} C_{\sigma \sigma'}({\bf K},{\bf Q}), \nonumber
\end{eqnarray}
that correspond to different ways of realizing the fluctuation
diagnostics. In fact, while $C$ is the quantity defined in the
original fluctuation diagnostics algorithm
where partial summations of all internal variables, but the transfer
momentum ${\bf Q}$, have been performed. $B$ corresponds to an equivalent scheme, where all internal variable are summed, except
the (incoming/outgoing) electron momentum $\bf{K'}$. This latter procedure can be,
thus viewed as a  ``fermionic''  reformulation of the fluctuation diagnostics. Finally, the
quantity $D$ is obtained performing the residual summation in the
definition of $B$ and $C$, and it depends, thus, only on the external
variables (here: on the momentum $\bf{K}$). 

By exploiting these definitions, we can rewrite Eq.~\ref{eq:dse_a} in
several equivalent ways:
\begin{eqnarray}\label{eq:11e}
\Sigma(k)+n_0U&=&{1 \over g(k)}\sum_{{\bf K}'{\bf Q}} A_{\uparrow\downarrow}({\bf K},{\bf K}',{\bf Q})\\
&=&{1 \over g(k)}\sum_{{\bf K}'} B_{\uparrow\downarrow}({\bf K},{\bf K}')\nonumber\\
&=&{1 \over g(k)}\sum_{{\bf Q}} C_{\uparrow\downarrow}({\bf K},{\bf Q})\nonumber\\
&=&{1 \over g(k)}D_{\uparrow\downarrow}({\bf K})+n_0U\nonumber. 
\end{eqnarray}
The Green's function given in Eq. (\ref{eq:1.1a}) can be written as
\begin{equation}\label{eq:11d}
g(k)= [i\nu_n-\Delta- \varepsilon_{\bf K}-\Sigma(k)]^{-1},
\end{equation}
Then we have 
\begin{equation}\label{eq:2.10c}
\Sigma(k)=[i\nu_n-\Delta- \varepsilon_{\bf K}]{D_{\uparrow\downarrow}({\bf K})\over 1+D_{\uparrow\downarrow}({\bf K})}.
\end{equation}
From this expression, we see immediately that $D \to -1$ leads to a
singularity, thus for $D\approx -1$ the system may become  a non Fermi
liquid or a Mott insulator. 

Performing the $\tau$ integration in Eq.~(\ref{eq:11a}), and
exploiting the Lehmann representation, we obtain
\begin{eqnarray}\label{eq:27}
&&A_{\sigma\sigma'}({\bf K},{\bf K}',{\bf Q})
={U\over N_cZ}  
\sum_{mnN}
{e^{-\beta \tilde E_m(N+1)} +e^{-\beta \tilde E_n(N)}\over i\nu +\tilde E_n(N)- \tilde E_m(N+1) }  \nonumber \\
&& \times \langle n|c_{{\bf K}+{\bf Q}\sigma}c^{\dagger}_{{\bf K}'+{\bf Q}\sigma'}c^{\phantom \dagger}_{{\bf K}'\sigma'}|m\rangle \langle m|
c_{{\bf K}\sigma}^{\dagger}|n\rangle,
\end{eqnarray}
where $\tilde E_n(N)=E_n(N)-\mu N$. 

\noindent

To simplify this expression, we now assume that $\beta$ is very large and that
\begin{eqnarray}\label{eq:15}
        &&\beta [\tilde E_1(N_{\rm el})-\tilde E_0(N_{\rm el})]\gg 1 \nonumber \\     
 &&\beta [\tilde E_1(N_{\rm el}\pm 1)-\tilde E_0(N_{\rm el}\pm 1)]\gg 1 \\   
 &&\beta [\tilde E_0(N_{\rm el}\pm 1)-\tilde E_0(N_{\rm el})]\gg 1,  \nonumber 
\end{eqnarray}
Here $N_{\rm el}$ is the number of electrons which minimizes $\tilde E_0(N)$.
For a very large and positive value of $U$ at half-filling, we have approximately that
$E_n(N_{\rm el}-1) O(U^0)$, $E_n(N_{\rm el})=O(U^0)$,    $E_n(N_{\rm el}+1)=U+O(U^0)$  
and $\mu=U/2$.  Similarly, for a very large, negative $U$ at half-filling, we have approximately that $E_n(N_{\rm el}-1)=(N_{\rm el}/2-1)U$,
$E_n(N_{\rm el})=(N_{\rm el}/2)U$, $E_n(N_{\rm el}+1)=(N_{\rm el}/2)U$
and $\mu=U/2$. 

In both cases, we obtain:
\begin{equation}\label{eq:16a}
\tilde E_n(N_{\rm el}\pm 1)-\tilde E_0(N_{\rm el})={|U| \over 2}+O(U^0). 
\end{equation}
Within the assumptions in Eq.~(\ref{eq:15}) we have, furthermore, that
\begin{equation}\label{eq:17}
Z=e^{-\beta \tilde E_0(N_{\rm el})},
\end{equation}
because all other contributions to $Z$ can be neglected and we can also neglect the
corresponding exponents in Eq.~(\ref{eq:11a}).
We have, thus, two types of contributions: i) $|n\rangle$ is the lowest ($N_{\rm el}$) state and
$|m\rangle$ is any ($N_{\rm el}+1$) state or ii) $|m\rangle$ is the lowest ($N_{\rm el}$) state
and $|n\rangle$ is any ($N_{\rm el}-1$) state.
As anticipated, we consider the lowest Matsubara frequency only, and assume that $\pi/\beta\ll |U|/2$, so that $\nu$ can be neglected.
Inserting Eq.~(\ref{eq:16a}) in Eq.~(\ref{eq:27}), we obtain
\begin{eqnarray}\label{eq:19b}
&&A_{\sigma\sigma'}({\bf K},{\bf K}',{\bf Q}) \nonumber \\
&&={4\over N_c}{U\over |U|}\langle E_0(N_{\rm el})|c_{{\bf K}\sigma}^{\dagger}
c^{\phantom \dagger}_{{\bf K}+{\bf Q}\sigma}c^{\dagger}_{{\bf K}'+{\bf Q}\sigma'}c^{\phantom \dagger}_{{\bf K}'\sigma'} |E_0(N_{\rm el})\rangle \nonumber \\
&&-{2\over N_c}{U\over |U|}\langle E_0(N_{\rm el})| c^{\dagger}_{{\bf K}'\sigma'}c^{\phantom \dagger}_{{\bf K}'\sigma'}|E_0(N_{\rm el})\rangle \delta_{{\bf Q}=(0,0)}  \\
&&-{2\over N_c}{U\over |U|}\langle E_0(N_{\rm el})|c_{{\bf K}+{\bf Q}\sigma}^{\phantom \dagger} c^{\dagger}_{{\bf K}+{\bf Q}\sigma} |E_0(N_{\rm el})\rangle \delta_{{\bf K}{\bf K}'}\delta_{\sigma \sigma'} \nonumber
\end{eqnarray} 

For the following discussion, it is useful to sum the first expectation value in 
Eq.~(\ref{eq:19b}) over ${\bf K}'$ and transformed to real space,
expressing $c_{{\bf K}\sigma}$ in the quantities $c_{{\bf R}\sigma}$.
\begin{eqnarray}\label{eq:19e}
&&\sum_{{\bf K}'}\langle E_0(N_{\rm el})|c_{{\bf K}\sigma}^{\dagger}c^{\phantom \dagger}_{{\bf K}+{\bf Q}\sigma}
c^{\dagger}_{{\bf K}'+{\bf Q}\sigma'}c^{\phantom \dagger}_{{\bf K}'\sigma'}
|E_0(N_{\rm el})\rangle \nonumber \\
&&={1\over N_c}\sum_{{\bf R}_1 {\bf R}_2{\bf R_3}}e^{i[{\bf K}\cdot ({\bf R}_1-{\bf R}_2)
+{\bf Q}\cdot ({\bf R}_3-{\bf R}_2 )]} \\ 
&& \times  \langle E_0(N_{\rm el})| c_{{\bf R}_1\sigma}^{\dagger}c^{\phantom \dagger}_{{\bf R}_2\sigma}
 c^{\dagger}_{{\bf R}_3\sigma'}c^{\phantom \dagger}_{{\bf R}_3\sigma'}  | E_0(N_{\rm el}) \rangle  \nonumber\\
&&\simeq{1\over N_c}\sum_{{\bf R}_1 {\bf R}_2}e^{i{\bf Q}\cdot ({\bf R}_2-{\bf R}_1 )} \nonumber \\ 
&& \times  \langle E_0(N_{\rm el})| n_{{\bf R}_1\sigma}
 n_{{\bf R}_2\sigma'}| E_0(N_{\rm el}) \rangle  \nonumber
\end{eqnarray}
Note that the last equation is obtained by neglecting the double
occupied/empty  (singly occupied) states, consistent with the
large positive (negative) interaction regime considered here.
These large $U$ approximations, the assumption in Eq.~(\ref{eq:16a})
and the neglect of double occupancy, have been discussed in Appendix 
\ref{sec:5.6}.

We now perform the ${\bf K}'$ and ${\bf Q}$ summations over $A_{\uparrow \downarrow}$.
\begin{eqnarray}\label{eq:19c}
&&D_{\uparrow \downarrow}({\bf K})+n_0Ug(k)=\sum_{{\bf K}'{\bf Q}}A_{\uparrow \downarrow}({\bf K},{\bf K}',{\bf Q})\nonumber \\
&&={U\over |U|}{4\over N_c}\sum_{{\bf R}_1}
\langle E_0(N_{el})|n_{{\bf R}_1\uparrow}n_{{\bf R}_1\downarrow}|E_0(N_{el})\rangle \nonumber\\
&&-{U\over |U|}{2\over N_c}\sum_{{\bf K}'}\langle E_0(N_{el})|n_{{\bf K}'\downarrow}|E_0(N_{el})\rangle 
\end{eqnarray}
For a half-filled system and a large positive $U$, the first term on the right
hand side contributes nothing, while the second terms contributes -1. For a large 
negative $U$ the first term instead contributes -2 and the second term +1. In both 
cases the total contribution to $D_{\uparrow \downarrow}({\bf K})=-1$. In the same limit the sum 
rule for $A_{\uparrow \uparrow}$ is zero. For the half-filled case and for ${\bf K}$ at the Fermi surface, the term 
$n_0Ug$ is zero to leading order in $1/U$, so that $D_{\uparrow \downarrow} \to -1$
and $D_{\uparrow \uparrow} \to 0$.

While the derivations of this section have been performed exploiting specific
assumptions, we will show later that  for large values of $U$ they agree with numerical results calculated 
{\sl without}  these assumptions (see also Appendix B). Thus they provide a reasonable theoretical framework for the physical
interpretion of our numerical data.

\subsection{Relation to localized ground states}\label{sec:3.3}

In Ref.~\onlinecite{Jaime} it has been argued that the progressive localization of some, and eventually  
of all, electrons on the embedded DCA cluster as $U$ is increased, plays an important 
role for the formation of the pseudogap. In particular, 
for large $U$, it was found that the electrons on the cluster localize 
into an RVB state. 

Below we discuss how the formulas [Eqs.~(\ref{eq:11}, 
\ref{eq:11a}, \ref{eq:11e}, \ref{eq:19b})] for the self-energy expressed 
in terms of the susceptibilities can be related to such a localization. 
In this subsection we assume that $U>0$.
To this aim, we now switch to working in real space. 
Performing the summation over ${\bf Q}$ in Eq.~(\ref{eq:19b}) we obtain
\begin{eqnarray}\label{eq:25}
&&B_{\uparrow \downarrow}({\bf K},{\bf K}',\nu)=\sum_{\bf Q}A_{\uparrow \downarrow}
({\bf K},{\bf K}',{\bf Q})=  \\
&&-{2\over N_c}\langle n_{{\bf K}'\downarrow} \rangle  
-{4\over N_c^2} \sum_{{\bf R}_1 \ne {\bf R}_2 } 
 e^{i({\bf K}-{\bf K}')\cdot ({\bf R}_1-{\bf R}_2)} \nonumber  \\
&&\times \langle E_0(N_{\rm el})|
c^{\dagger}_{{\bf R}_2\downarrow} c^{\phantom\dagger}_{{\bf R}_2\uparrow}c_{{\bf R}_1\uparrow}^{\dagger}
c^{\phantom \dagger}_{{\bf R}_1\downarrow}|E_0(N_{\rm el})\rangle.  \nonumber
\end{eqnarray}
Here we have introduced similar approximations as in Eq.~(\ref{eq:19e}), assuming
half-filling and  a large $U$.

The operator in Eq.~(\ref{eq:25}) couples to a valence bond
\begin{equation}\label{eq:27a}
(12)={1\over \sqrt{2}} (c^{\dagger}_{{\bf R}_1\uparrow}c^{\dagger}_{{\bf R}_2\downarrow}
-c^{\dagger}_{{\bf R}_1\downarrow}c^{\dagger}_{{\bf R}_2\uparrow})|{\rm vac}\rangle
\end{equation}
between the sites ${\bf R}_1$ and ${\bf R}_2$ with the strength -$1/2$. Here $|{\rm vac}\rangle$
is the vacuum state.  For a nearest
neighbor (NN) bond on a square lattice and ${\bf K}-{\bf K}'=(\pi,\pi)$  we then obtain a
negative contribution due to the sign of the exponential, while for ${\bf K}-{\bf K}'=(0,0)$
the contribution is positive. 

For a square lattice we can divide the lattice in two sublattices A and B, where the
nearest neighbor of a site in one sublattice belongs to the other sublattice.
The NN-RVB is a superposition of singlets of the type of Eq.~(\ref{eq:27a})
between neighboring sites taken from A to B
sublattice with equal positive bond amplitudes:
\begin{equation}
| \Psi_0 \rangle =\sum_{i_\alpha,j_\beta} (i_1j_1)(i_2j_2)...(i_nj_n)
\label{RVB}
\end{equation}
where $i_\alpha$ ($j_\beta$) denote neighbor sites in the A-sublattice (B-sublattice) 
and $(i_\alpha$ $ j_\beta)$ denotes a singlet. First we only consider contributions 
where the operator in Eq.~(\ref{eq:25}) acts within a given bond. For a NN-RVB state these contributions together
with the term $-2\left<n_{{\bf K}' \downarrow}\right>/N_c$ gives a substantial negative result      
for ${\bf K}-{\bf K}'=(\pi,\pi)$, while there tends to be a net positive contribution 
for ${\bf K}={\bf K}'$. In addition there are also contributions where the operator 
in Eq.~(\ref{eq:25}) acts on a product of two bonds $(i_1j_1)(i_2j_2)$
, i.e., the two-site operator acts on one site of each bond. Then we 
can get a contribution also from second nearest neighbor, even for a NN-RVB state.
In this case the exponent in Eq.~(\ref{eq:25}) takes the same value for both
${\bf K}-{\bf K}'=(\pi,\pi)$ and ${\bf K}-{\bf K}'=(0,0)$, reducing the difference between
these two cases. As we will show in the following,
$B_{\uparrow\downarrow}({\bf K},{\bf K}')$ encodes, nevertheless, very
clear signals of the RVB state formation. 

Within the approximations above, the large $U$ results (\ref{eq:25}) only depends on 
${\bf K}-{\bf K}'$ and not on the two wave-vectors individually. For finite values 
of $U$, however, there is also a difference between ${\bf K}=(\pi,0)$ and
${\bf K}=(\pi/2,\pi/2)$, which plays an important role in the interpretation 
of the results.

\subsection{Relation to the ``fluctuation diagnostics''}\label{sec:3.4}
We now discuss the relation to ``fluctuation diagnostics'' \cite{Fluct} introduced
earlier. While for the SU(2)-symmetric case, this connection can be
analytically derived from the exact relations holding between the
correlation functions\cite{su2note} 
$B$ and $C$. Here we however 
provide a physically more explicit derivation. 
To this aim, we first consider the coupling to charge and spin fluctuations. 
For this purpose we introduce 
\begin{eqnarray}\label{eq:19f}
C_{\rm ch/ sp}({\bf K},{\bf Q})=\sum_{{\bf K}'}[A_{\uparrow\downarrow}({\bf K},{\bf K}',{\bf Q})\pm
A_{\uparrow\uparrow}({\bf K},{\bf K}',{\bf Q})],
\end{eqnarray}
and 
\begin{equation}\label{eq:19l}
\Sigma(k)=\sum_{\bf Q}C_{ch/sp}({\bf K},{\bf Q})/g(k).
\end{equation}
Here we have used that $\sum_{{\bf K}'{\bf Q}}A_{\uparrow\uparrow}({\bf K},{\bf K}',{\bf Q})=0$.

The coupling to spin fluctuations is expected to be relevant for
positive values of $U$. In particular, for large enough $U$, at half-filling,
we have:
\begin{eqnarray}\label{eq:19g}
&&C_{\rm sp}({\bf K},{\bf Q})=
-{2\over N_c^2}\sum_{{\bf R}_1 {\bf R}_2}e^{i{\bf Q}\cdot ({\bf R}_2-{\bf R}_1 )}  \\
&&\times \langle E_0(N_{\rm el})| [n_{{\bf R}_1\uparrow}-n_{{\bf R}_1\downarrow}]  
 [n_{{\bf R}_2\uparrow} 
-n_{{\bf R}_2\downarrow}]| E_0(N_{\rm el}) \rangle  \nonumber \\
&&+{2\over N_c}\langle E_0(N_{\rm el})|c_{{\bf K}+{\bf Q}\uparrow}^{\phantom \dagger} c^{ \dagger}_{{\bf K}+{\bf Q}\uparrow} |E_0(N_{\rm el})\rangle. \nonumber
\end{eqnarray}
One can then see that if neighboring sites have opposite spins 
a large (negative) contribution to $C_{\rm sp}$ is obtained for ${\bf
  Q}=(\pi,\pi)$. This is indeed perfectly consistent with the
considerations made in the
previous subsection due to an exact relation which can be derived in the SU(2)-symmetric case:
$C_{\rm sp} ({\bf K}, {\bf Q}=(\pi,\pi)) =B_{\uparrow \downarrow}({\bf K}, {\bf K}+(\pi,\pi))$.\\
We next consider the coupling to charge fluctuations. These are 
expected to become important for negative $U$, and therefore we 
consider $U<0$ below. In Eq.~(\ref{eq:19b}), $U/|U|$ then contributes 
an extra minus sign. Eq.~(\ref{eq:19e}) is still a good approximation 
for large $U$, but now because double occupancy is favored.
We then have
\begin{eqnarray}\label{eq:19h}
&&C_{\rm ch}({\bf K},{\bf Q})=
	-{2\over N_c^2}\sum_{{\bf R}_1 {\bf R}_2}e^{i{\bf Q}\cdot ({\bf R}_2-{\bf R}_1 )}\langle E_0(N_{\rm el})|  \\
&&\times [n_{{\bf R}_1\uparrow}+n_{{\bf R}_1\downarrow}-\langle n \rangle]
[n_{{\bf R}_2\uparrow} 
+n_{{\bf R}_2\downarrow}-\langle n \rangle]| E_0(N_{\rm el}) \rangle  \nonumber \\
&&+{2 \over N_c}\langle E_0(N_{\rm el})|c_{{\bf K}+{\bf Q}\uparrow}^{\phantom \dagger} c^{ \dagger}_{{\bf K}+{\bf Q}\uparrow} |E_0(N_{\rm el}) \rangle\nonumber \\
&&+2 (\langle n \rangle -\langle n\rangle^2))\delta_{{\bf Q}(0,0)}, \nonumber
\end{eqnarray}
where we have assumed that there is no net spin polarization.
If there is a checkerboard type of charge order, there is a 
large negative contribution for ${\bf Q}=(\pi,\pi)$.

While Eqs.~(\ref{eq:2.10}) makes contact to susceptibilities 
via Eq.~(\ref{eq:2.3}), Eqs.~(\ref{eq:19g}, \ref{eq:19h}) 
make direct contact to static charge or spin correlations 
in real space.  This connection is based on the approximation
(\ref{eq:16a}). 

Finally, we consider the coupling to superconductivity fluctuations. 
As discussed in Ref.~\onlinecite{Fluct}, the coupling to unconventional 
superconductivity is weak, and the coupling is primarily to $s$-wave 
superconductivity. In the Hubbard model, this is found for negative $U$. 
We then consider $U<0$.
To study superconductivity we define\cite{Rohringer}
\begin{eqnarray}\label{eq:19i}
&&C_{\rm pp}({\bf K},{\bf Q})=\sum_{{\bf K}'}A_{\uparrow \downarrow}({\bf K},{\bf K}',{\bf Q}-{\bf K}-{\bf K}') \\
&&=-{4\over N_c^2}\sum_{{\bf R}_1 {\bf R}_2 {\bf R}_3}e^{i[{\bf K}\cdot({\bf R}_1-{\bf R}_3)-{\bf Q}\cdot({\bf R}_2-{\bf R}_3)]}  \nonumber \\
&&\langle E_0(N_{\rm el})|c^{\dagger}_{{\bf R}_1\uparrow}c^{\dagger}_{{\bf R}_3\downarrow}
c^{\phantom \dagger}_{{\bf R}_2\downarrow}c^{\phantom \dagger}_{{\bf R}_2\uparrow}| E_0(N_{\rm el}) \rangle \nonumber \\
&&+\langle n \rangle  \delta_{{\bf Q} (0,0)} \nonumber
\end{eqnarray}
We are interested in the ${\bf Q}=(0,0)$ contribution. The contribution 
in the sum in Eq.~(\ref{eq:19i}) for ${\bf R}_3={\bf R}_1$ is then 
particularly important, since the exponential function is then unity.
The coupling to $s$-wave superconductivity is then strong.

Using Eq.~(\ref{eq:11c}), we can see that a spin density wave contributes to $D\to -1$ in Eq.~(\ref{eq:19g}).
The same is true for a charge density wave in Eq.~(\ref{eq:19h}) and a 
$s$-wave superconductivity pairing in Eq.~(\ref{eq:19i}). In all three cases this 
contributes to the opening of a gap.

\section{RVB state for isolated clusters}\label{sec:4}

In this section, we study the different manifestations of the onset of
a RVB-like ground state in the correlation functions.
It is known that an RVB state\cite{Anderson} is formed for small isolated clusters
for large values of $U$.\cite{Tang,Jaime}
We will start by considering the four-level model, where there are two ${\bf K}$-states on the cluster, 
$(\pi,0)$ and $(0,\pi)$, each coupling to one bath state. This model shows the relations
between the real space correlations and the correlation function $B$ in a particularly transparent way. 
We then extend our analysis to larger isolated clusters, where the
physics is similar.  Eventually, the general connection to the spectral properties of
the systems will be discussed in Sec.~\ref{sec:5}.

\subsection{Four-level model}\label{sec:4.1}

In the limit of very large $U$ and for $\Delta U>0$, the ground-state 
of the four-level model takes the form
\begin{equation}\label{eq:5.5}
|\Phi\rangle={1\over 2}(c^{\dagger}_{1c\uparrow}c^{\dagger}_{1c\downarrow}-
c^{\dagger}_{2c\uparrow}c^{\dagger}_{2c\downarrow})
(c^{\dagger}_{1b\uparrow}c^{\dagger}_{1b\downarrow}-
c^{\dagger}_{2b\uparrow}c^{\dagger}_{2b\downarrow})|{\rm vac}\rangle,
\end{equation}
where the indices 1 and 2 refer to orbitals with ${\bf K}=(\pi,0)$ and $(0,\pi)$, respectively.
The ground state wave function is a product of a cluster and a bath wave function, 
and the hopping between the cluster and the bath is unimportant in this limit. In this sense we can 
talk about a localization of two electrons on the cluster.  Writing explicitly the cluster
part of the wave function we get
\begin{equation}\label{eq:5.2a}
|{\rm \Phi}_{\rm c}\rangle \!= \!{1\over \sqrt{2}}(c^{\dagger}_{(\pi,0)\uparrow}c^{\dagger}_{(\pi,0)\downarrow}-
c^{\dagger}_{(0,\pi)\uparrow}c^{\dagger}_{(0,\pi)\downarrow})|{\rm vac}\rangle.
\end{equation}
We also note, that  the corresponding state for a four-site cluster with the ${\bf K}=(0,0)$ orbital
occupied is given by
\begin{equation}\label{eq:5.2}
|{\rm \psi}_{\rm loc}\rangle \!= \!{1\over \sqrt{2}}(c^{\dagger}_{(\pi,0)\uparrow}c^{\dagger}_{(\pi,0)\downarrow}-
c^{\dagger}_{(0,\pi)\uparrow}c^{\dagger}_{(0,\pi)\downarrow})c^{\dagger}_{(0,0)\uparrow}
c^{\dagger}_{(0,0)\downarrow}|{\rm vac}\rangle,
\end{equation}
In Ref.~\onlinecite{Parquetprb} it was shown that this is a good approximation to the 
calculated state for a four-site cluster for moderate values of $U$. It was also shown
 that this state is fairly closely related to a RVB state for the four-site cluster. 
We therefore refer to the states in Eqs.~(\ref{eq:5.2a}, \ref{eq:5.2}) as RVB-like states, 
although the RVB state is not yet fully developed for values of $U$ of interest here. 
This is discussed further in Appendix \ref{sec:5.6}. The states have
singlet character. 

If, for large $U$, $\Delta U<0$ instead, the ground-state of the 
four-level model takes the form
\begin{eqnarray}\label{eq:5.6}
&&|\Phi\rangle={1\over \sqrt{3}}[c^{\dagger}_{1c\uparrow}c^{\dagger}_{2c\uparrow}
c^{\dagger}_{1b\downarrow}c^{\dagger}_{2b\downarrow} 
-{1\over 2}(c^{\dagger}_{1c\uparrow}c^{\dagger}_{2c\downarrow}+c^{\dagger}_{1c\downarrow}c^{\dagger}_{2c\uparrow})\nonumber \\
&&\times (c^{\dagger}_{1b\uparrow}c^{\dagger}_{2b\downarrow}+c^{\dagger}_{1b\downarrow}c^{\dagger}_{2b\uparrow})
+c^{\dagger}_{1c\downarrow}c^{\dagger}_{2c\downarrow}
c^{\dagger}_{1b\uparrow}c^{\dagger}_{2b\uparrow}]|{\rm vac}\rangle.
\end{eqnarray}
Here the $c$ part is degenerate and has triplet character and it couples to the bath in a Kondo-like way.
In this case, even in the large $U$  limit we cannot separate the wave function as a product of a cluster and a bath part,
and hopping remains essential for the character of the state. We can then not talk about localization 
in the sense used below Eq.~(\ref{eq:5.5}). 

In order to clarify the connection with the correlations in real
space,  we first convert the $(\pi,0)$ and $(0,\pi)$ states to real space,
using a four site cluster to represent the levels. Putting sites 1 and 2 in the $x$-direction 
and 1 and 4 in the $y$-direction (see inset in Fig.~\ref{fig:1}), we have 
\begin{eqnarray}\label{eq:71}
&&\phi_{(\pi,0)}={1\over 2}(|1\rangle-|2\rangle+|4\rangle-|3\rangle) \\
&&\phi_{(0,\pi)}={1\over 2}(|1\rangle+|2\rangle-|4\rangle-|3\rangle)   \nonumber
\end{eqnarray}
The wave function in Eq.~(\ref{eq:5.2a}) then takes the (real space) form
\begin{eqnarray}\label{eq:72}
&& | E_0(N_{\rm el}) \rangle={1\over 2\sqrt{2}}[-(|1\uparrow 2\downarrow\rangle -| 1\downarrow2\uparrow\rangle)
+(|1\uparrow 4\downarrow\rangle \\
&&-| 1\downarrow4\uparrow\rangle) 
+(|2\uparrow 3\downarrow\rangle -| 2\downarrow3\uparrow\rangle)
-(|3\uparrow 4\downarrow\rangle -| 3\downarrow4\uparrow\rangle)]\nonumber.
\end{eqnarray}
The two terms within each parenthesis $(...)$ form a valence bond (VB), and 
the sum over all nearest neighbor VB gives the nearest neighbor RVB form.
It is then clear that the operator in Eq.~(\ref{eq:25}) couples the terms 
{\sl within} each parenthesis very efficiently. The different contributions to
$B_{\uparrow \downarrow}$ add up coherently (with a negative sign) for ${\bf K}-{\bf K}'=(\pi,\pi)$ 
and combine constructively with the trivial term 
$\langle n_{{\bf K}'\downarrow}\rangle$, while for ${\bf K}-{\bf K}'=0$ 
there is a destructive interference with term $\langle n_{{\bf K}'\downarrow}\rangle$.  
The result is a large negative contribution for ${\bf K}-{\bf K}'=(\pi,\pi)$ 
and a small result for ${\bf K}-{\bf K}'=0$.  The results for $B_{\uparrow 
\downarrow}({\bf K},{\bf K}')$ are shown in Table~\ref{table:2}. The corresponding results 
for $A_{\uparrow \downarrow}({\bf K}, {\bf K}',{\bf Q})$ and $C_{\rm sp}({\bf K},
{\bf Q})$ are also reported there: The main contribution to 
$B_{\uparrow \downarrow}[{\bf K},{\bf K}+(\pi,\pi)]$ originates from $A_{\uparrow 
\downarrow}[{\bf K},{\bf K}+(\pi,\pi),(\pi,\pi)]$. The same holds for  $C_{\rm sp}({\bf K},{\bf Q})$ at the corresponding
transfer momentum ${\bf Q}=(\pi,\pi)$. 

The large value of $A_{\uparrow \downarrow}[(\pi,0);(0,\pi);(\pi,\pi)]$ can also 
be easily understood in reciprocal space by applying the definition of $A$ in
Eq.~(\ref{eq:11a}) to the wave function in Eq.~(\ref{eq:5.2a}), because the operator 
in Eq.~(\ref{eq:11a}) directly couples the two terms in Eq.~(\ref{eq:5.2a}). 

For $\Delta U<0$ the behavior is completely different, as seen in Table~\ref{table:2a}. 
$|B_{\uparrow \downarrow}({\bf K},{\bf K}')|$ is now particularly large for ${\bf K}'={\bf K}$,
and the main contribution to this $B$ comes from ${\bf Q}=(0,0)$. At
the same time, $C_{\rm sp}({\bf K},{\bf Q})$ is now the same for ${\bf Q}=(0,0)$ and
$(\pi,\pi)$. Further, in contrast to almost of all other cases we
consider in this work, we note that here $C_{\rm sp}({\bf K},{\bf Q}= (\pi,\pi ))$ differs significantly from 
$B_{\uparrow \downarrow}({\bf K},{\bf K}+ (\pi,\pi))$. This happens
because, as we already mentioned in Sec.~\ref{sec:2}, in the four-level model the SU(2) symmetry is
violated and, thus, the related equivalence relations between $B$ and
$C$ are not guaranteed a priori.\cite{noteSU2vio}
\begin{table}
\caption{\label{table:2}Re $A_{\uparrow \downarrow}({\bf K},{\bf K}',{\bf Q})$ for the four-level 
model in the large $U$ and $T\to 0$ limits and for $\Delta U>0$. ${\bf K}=(\pi,0)$.
The table also shows Re $B_{\uparrow \downarrow}({\bf K},{\bf K}')$ and Re $C_{\rm sp}({\bf K},{\bf Q})$.
}
\begin{tabular}{crrrrrr}
\hline
\hline
${\bf Q}$ & \multicolumn{1}{c}{${\bf K}'={\bf K}+(\pi,\pi)$} & \multicolumn{1}{c}{${\bf K}'={\bf K}$} & $C_{\rm sp}({\bf K},{\bf Q})$\\
\hline
$(\pi,\pi)$ & -1.0& 0.0 & -1.5\\
$(0,0)$      &-0.5& 0.5 & 0.5 \\
\hline
$B_{\uparrow \downarrow}({\bf K},{\bf K}')$       & -1.5 & 0.5 & \\
\hline
\hline
\end{tabular}
\end{table}

\begin{table}
\caption{\label{table:2a}Re $A_{\uparrow \downarrow}({\bf K},{\bf K}',{\bf Q})$ for the four-level 
model in the large $U$ and $T\to 0$ limits and for $\Delta U<0$. ${\bf K}=(\pi,0)$.
The table also shows Re $B_{\uparrow \downarrow}({\bf K},{\bf K}')$ and Re $C_{\rm sp}({\bf K},{\bf Q})$.
}
\begin{tabular}{crrrrrr}
\hline
\hline
${\bf Q}$ & \multicolumn{1}{c}{${\bf K}'={\bf K}+(\pi,\pi)$} & \multicolumn{1}{c}{${\bf K}'={\bf K}$} & $C_{\rm sp}({\bf K},{\bf Q})$\\
\hline
$(\pi,\pi)$ & 0.00& -0.33 & -0.5 \\
$(0,0)$      &-0.16& -0.50 & -0.5  \\
\hline
$B_{\uparrow \downarrow}({\bf K},{\bf K}')$       & -0.16 & -0.83 & \\
\hline
\hline
\end{tabular}
\end{table}

\begin{table}
\caption{\label{table:1}Re $B_{\uparrow \downarrow}({\bf K},{\bf K}')$ at $T=0$ for an isolated cluster with $N_c$ sites
in the large $U$ limit according to Eq.~(\ref{eq:25}). 
}
\begin{tabular}{rcc}
\hline
\hline
Model & ${\bf K}'={\bf K}+(\pi,\pi)$ & ${\bf K}'={\bf K}$   \\
\hline
Four-level ($N_c=2)$   & -1.5 & 0.50 \\
$N_c=4$     &-1.09 & 0.25 \\
$N_c=8$     & -0.88 & 0.13 \\
$N_c=16$   & -0.51 & 0.06 \\
\hline 
\hline
\end{tabular}
\end{table}

\subsection{Isolated $N_c=4$ cluster}\label{sec:4.3}

We next consider the $N_c=4$ case in the large $U$ limit. The wave function of the cluster takes
the RVB form, which can be written as
\begin{eqnarray}\label{eq:29}
&&\psi={1\over \sqrt{3}}
[c^{\dagger}_{{\bf R}_1\uparrow}c^{\dagger}_{{\bf R}_2\downarrow}
c^{\dagger}_{{\bf R}_3\uparrow}c^{\dagger}_{{\bf R}_4\downarrow}
+c^{\dagger}_{{\bf R}_1\downarrow}c^{\dagger}_{{\bf R}_2\uparrow}
c^{\dagger}_{{\bf R}_3\downarrow}c^{\dagger}_{{\bf R}_4\uparrow}] |{\rm vac}\rangle \nonumber \\
&&-{1\over 2\sqrt{3}}[
c^{\dagger}_{{\bf R}_1\uparrow}c^{\dagger}_{{\bf R}_2\downarrow}
c^{\dagger}_{{\bf R}_3\downarrow}c^{\dagger}_{{\bf R}_4\uparrow}
+c^{\dagger}_{{\bf R}_1\downarrow}c^{\dagger}_{{\bf R}_2\uparrow}
c^{\dagger}_{{\bf R}_3\uparrow}c^{\dagger}_{{\bf R}_4\downarrow}  \\
&&+c^{\dagger}_{{\bf R}_1\uparrow}c^{\dagger}_{{\bf R}_2\uparrow}
c^{\dagger}_{{\bf R}_3\downarrow}c^{\dagger}_{{\bf R}_4\downarrow}
+c^{\dagger}_{{\bf R}_1\downarrow}c^{\dagger}_{{\bf R}_2\downarrow}
c^{\dagger}_{{\bf R}_3\uparrow}c^{\dagger}_{{\bf R}_4\uparrow}]|{\rm vac}\rangle  \nonumber
\end{eqnarray}

Here we will focus again on the most relevant case, where ${\bf K}=(\pi,0)$ and ${\bf K}'=(\pi,0)$ or $(0,\pi)$.
Applying the operator in  Eq.~(\ref{eq:25}), leads to couplings involving NN bonds          
and connecting terms between the first row  and the second or third rows. Using 
the large $U$ limit [Eq.~(\ref{eq:16a})], we obtain $B_{\uparrow \downarrow}({\bf K},{\bf K}')=-0.92$ 
for ${\bf K}'={\bf K} +(\pi,\pi)$ and $+0.42$ for ${\bf K}'={\bf K}$. Although the wave function in Eq.~(\ref{eq:29}) has no
second nearest neighbor RVB, couplings are generated between the terms
in the second and third rows, in particular second 
nearest-neighbor couplings. The function $B_{\uparrow \downarrow}[{\bf
  K},{\bf K}+(\pi,\pi)]$ s therefore not a perfect
measure of the RVB nature of the ground-state. These terms, however, are rather small ($0.17$), and the 
total contributions to $B_{\uparrow \downarrow}[{\bf
  K},{\bf K'}]$ change from $-0.92$ to $-1.09$ and from $+0.42$ to $+0.25$, 
respectively (cf. Table~\ref{table:1}).  

To double-check the reliability of our approximations, we have also calculated results 
for $A_{\uparrow \downarrow}({\bf K},{\bf K}',{\bf Q})$ 
for a $N_c=4$ isolated cluster at $T=0$ for a finite $U=1.6$ eV and for $U=\infty$ without 
using any of the assumptions behind Eq.~(\ref{eq:25}). The states of the isolated cluster 
were calculated using exact diagonalization, and expressed in terms of orbitals with 
well-defined ${\bf K}$-values. Eq.~(\ref{eq:27}) can then easily be applied directly. 
Table ~\ref{table:3} compares the results with the approximate formula Eq.~(\ref{eq:25}). 
The approximate results are generally smaller (in absolute values) than the exact results 
for $U=1.6$ eV. This is primarily due to the approximation in Eq.~(\ref{eq:16a}) for 
the eigenvalues, which leads to an underestimate in Table~\ref{table:3} by a factor 0.6 
for $U=1.6$ eV.  Taking this into account, the agreement between the approximate and the 
exact results is fairly good.  For $U=\infty$, Eq.~(\ref{eq:25}) becomes exact, and gives
the exact results shown in the table. The table also shows the
corresponding numerical results for $C_{\rm sp}$,
demonstrate that this quantity also becomes substantially large (and negative) for
${\bf Q}=(\pi,\pi)$, consistent with its equivalence to $B_{\uparrow \downarrow}[{\bf
  K},{\bf K}+(\pi,\pi)]$  in SU(2)-symmetric calculations. 

For $N_c=4$ and $U=1.6$ eV, the ${\bf K}=(0,0)$ level is approximately doubly occupied, 
and the remaining two electrons form a wave function similar to the cluster part of 
Eq.~(\ref{eq:5.2a}) for the four-level model,\cite{Parquetprb} as shown in Eq.~(\ref{eq:5.2}). 
It is then natural that the results are similar to the ones for the
four-level model.\cite{notelevels} 
In particular, there is an important contribution to $B_{\uparrow \downarrow}[{\bf K},
{\bf K}+(\pi,\pi)]$ from ${\bf Q}=(\pi,\pi)$. For $U=\infty$ all levels are occupied
equally, and Eq.~(\ref{eq:5.2}) is not a good approximation any more. Then the four-site
model becomes quite different from the four-level model.

\begin{table}
\caption{\label{table:3}Re $A_{\uparrow \downarrow}({\bf K},{\bf K}',{\bf Q})$ for an $N_c=4$ isolated cluster
and ${\bf K}=(\pi,0)$. Exact results [Eq.~(\ref{eq:27})] and results obtained using the 
approximate formula in Eqs.~(\ref{eq:25}) are shown for $U=1.6$ eV. For $U=\infty$,
Eq.~(\ref{eq:25}) becomes exact, and therefore only exact results are shown.
The parameters are $t=-0.50$ are $\beta=\infty$. The corresponding
values of $C_{\rm sp}$ have been computed for the case $U=1.6$ eV and $\beta=60$ eV$^{-1}$.
}

\begin{tabular}{rrrrrrrr}
\hline
\hline
${\bf Q}$ & \multicolumn{3}{c}{${\bf K}'={\bf K}+(\pi,\pi)$} & \multicolumn{3}{c}{${\bf K}'={\bf K}$} & $C_{\rm sp}({\bf Q})$ \\
& \multicolumn{2}{c}{$U=1.6$ eV}     & \multicolumn{1}{c}{$U$=$\infty$}    &\multicolumn{2}{c}{$U=1.6$ eV}   & \multicolumn{1}{c}{$U$=$\infty$}    \\
& Ex & Appr. & Ex &  Ex & Appr. &Ex&  \\
\hline
$(\pi,\pi)$ & -0.67&-0.45 & -0.38 &  -0.02&-0.01 &-0.04 & -1.35   \\
$(0,0)$      &-0.35&-0.23  &-0.21 &   0.31& 0.20  & 0.12 & 0.37  \\
$(\pi,0)$ & -0.17&-0.13 &-0.25 &  0.03 &0.03 & 0.08 & 0.01 \\
$(0,\pi)$ & -0.17&-0.13 & -0.25&   0.03 &0.03 & 0.08 & 0.01 \\
\hline
$B({\bf K},{\bf K}')_{\uparrow \downarrow}$       & -1.36&-0.95 & -1.09&   0.35 &0.25 & 0.25 &  \\
\hline
\hline
\end{tabular}
\end{table}

\subsection{Larger isolated clusters}\label{sec:4.4}

The dependences of our isolated cluster results on the cluster sizes
have also been studied systematically, and the corresponding results
have been summarized in Table~\ref{table:1}. The calculations have been performed 
using the approximate formula in Eq.~(\ref{eq:25})
for isolated clusters of size $N_c$ in the large $U$ limit. More specifically
Eq.~(\ref{eq:16a})  has been used for evaluating the energy differences, and the results should therefore
be exact in the large $U$ limit. We also recall that for $N_c=4$ and $8$, the exact ground-state is a NN-RVB 
state, while for $N_c=16$ also more distant bonds (neglected in our calculations) play a role.

From the results of Table~\ref{table:1}, we can verify, once again, how the RVB character of the states is signaled by the large negative value of 
$B({\bf K},{\bf K}')$ for ${\bf K}-{\bf K}'=(\pi,\pi)$  and a smaller positive value for 
${\bf K}={\bf K}'$.

\section{Relation to spectra}\label{sec:5}

In this section, we eventually discuss one of the central topics of
this work:  the relation between real space correlations 
and spectral functions. Our purpose is to illustrate how changes in these
correlation functions are reflected in the spectral function, often in a dramatic way.
This will make it possible, in turn, to use spectral functions to extract information about
correlation functions. In particular, we will focus here on RVB-like correlations.

In Sec.~\ref{sec:3}, we have already established relations between real space correlation functions 
and the behavior of the self-energy $\Sigma$ for small $\nu_n$.
The behavior
of $\Sigma$ controls, in turn, the behavior of the spectrum. We have that
\begin{eqnarray}\label{eq:29a}
&&{\rm Im}\ g(k)={\rm Im} {1\over i\nu_n-\Delta -\bar \varepsilon_{\bf K}-\Sigma(k)} \\
&&=-\nu_n\int {\rho({\bf k},\varepsilon) \over \varepsilon^2+\nu_n^2} d\varepsilon, \nonumber
\end{eqnarray}
i.e. Im $g(k)$ measures the spectral weight over a range $\nu_n$ around $\varepsilon=0$. 
It is then clear that the presence of large negative value of Im $\Sigma$ 
for small $\nu_n$ will be reflected in a small or vanishing $\rho$ for small $\varepsilon$,
i.e. the system behaves as a non Fermi liquid or, even, as  a Mott-Hubbard insulator.  

\subsection{Four-level model}\label{sec:5.1}

\begin{figure}
{\rotatebox{-90}{\resizebox{6.0cm}{!}{\includegraphics {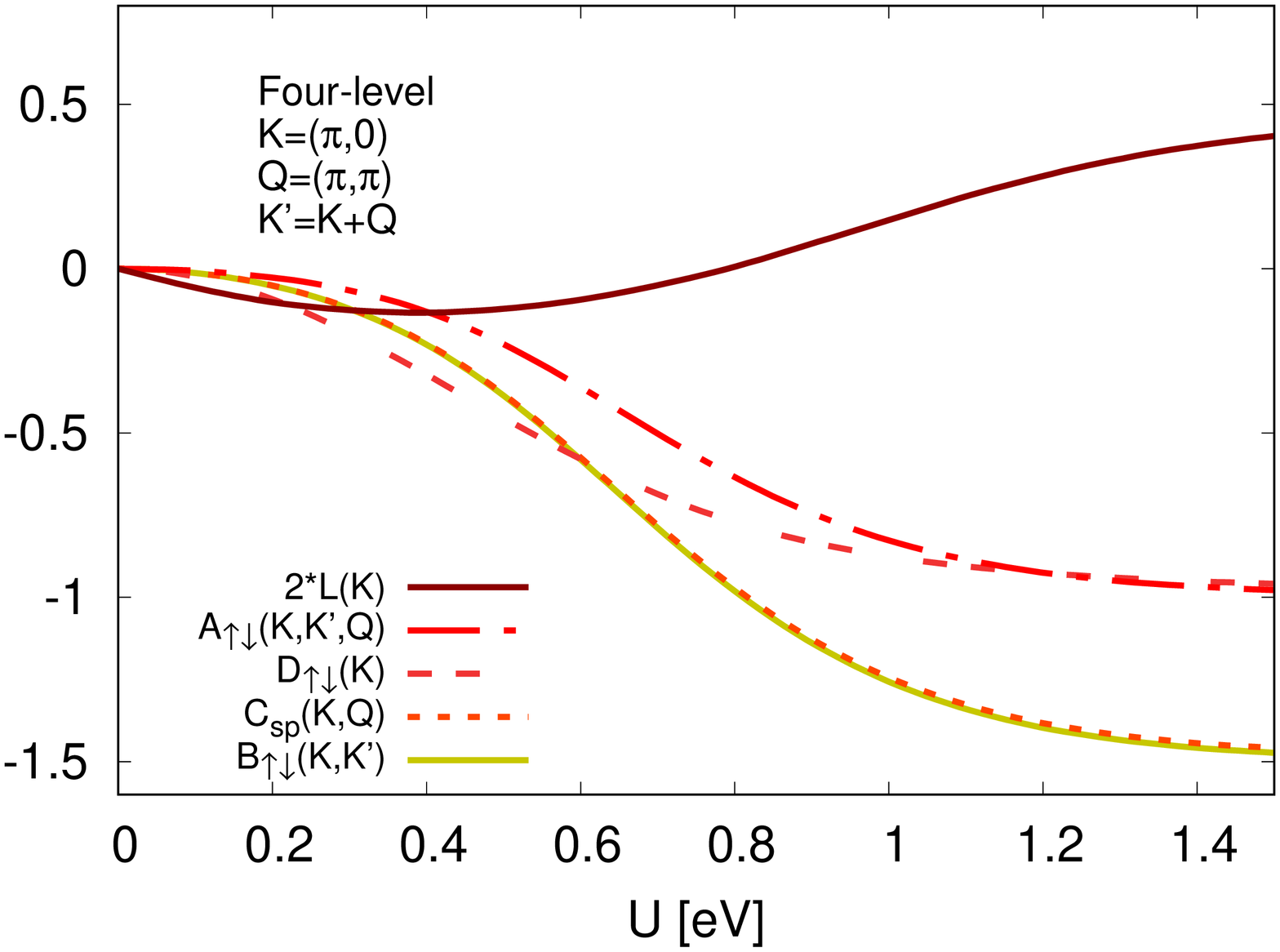}}}}
\vskip-4.2cm
\hskip4.5cm
{\rotatebox{0}{\resizebox{1.5cm}{!}{\includegraphics {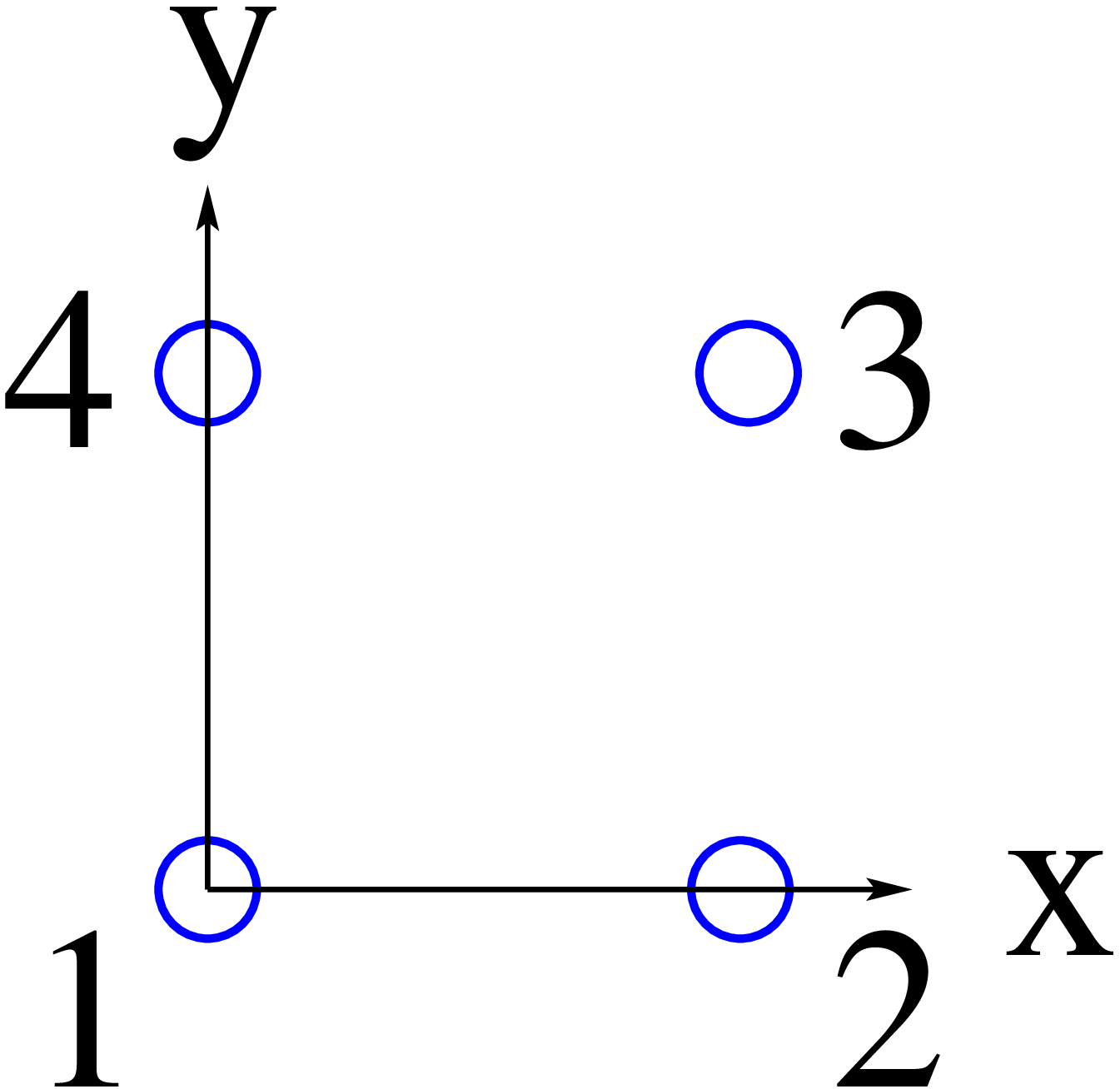}}}}
\vskip4.5cm
\caption{
Real parts of $D_{\uparrow \downarrow} ({\bf K})$, $B_{\uparrow \downarrow}({\bf K},
{\bf K}')$, $C_{\rm sp}({\bf K},{\bf Q})$ and $A_{\uparrow \downarrow}({\bf K},{\bf K}' ,{\bf Q})$
of the four-level model for ${\bf K}=(\pi,0)$,  ${\bf K}' =(0,\pi)$ and ${\bf Q}=(\pi,\pi)$ 
as well as $L({\bf K})$. The inset shows the numbering of the sites in Eq.~(\ref{eq:71}).
The parameters are  $V=-0.1$ eV, $\Delta U>0$ and $\beta=60$ eV$^{-1}$.
}\label{fig:1}
\end{figure}

\begin{figure}
{\rotatebox{-90}{\resizebox{6.0cm}{!}{\includegraphics {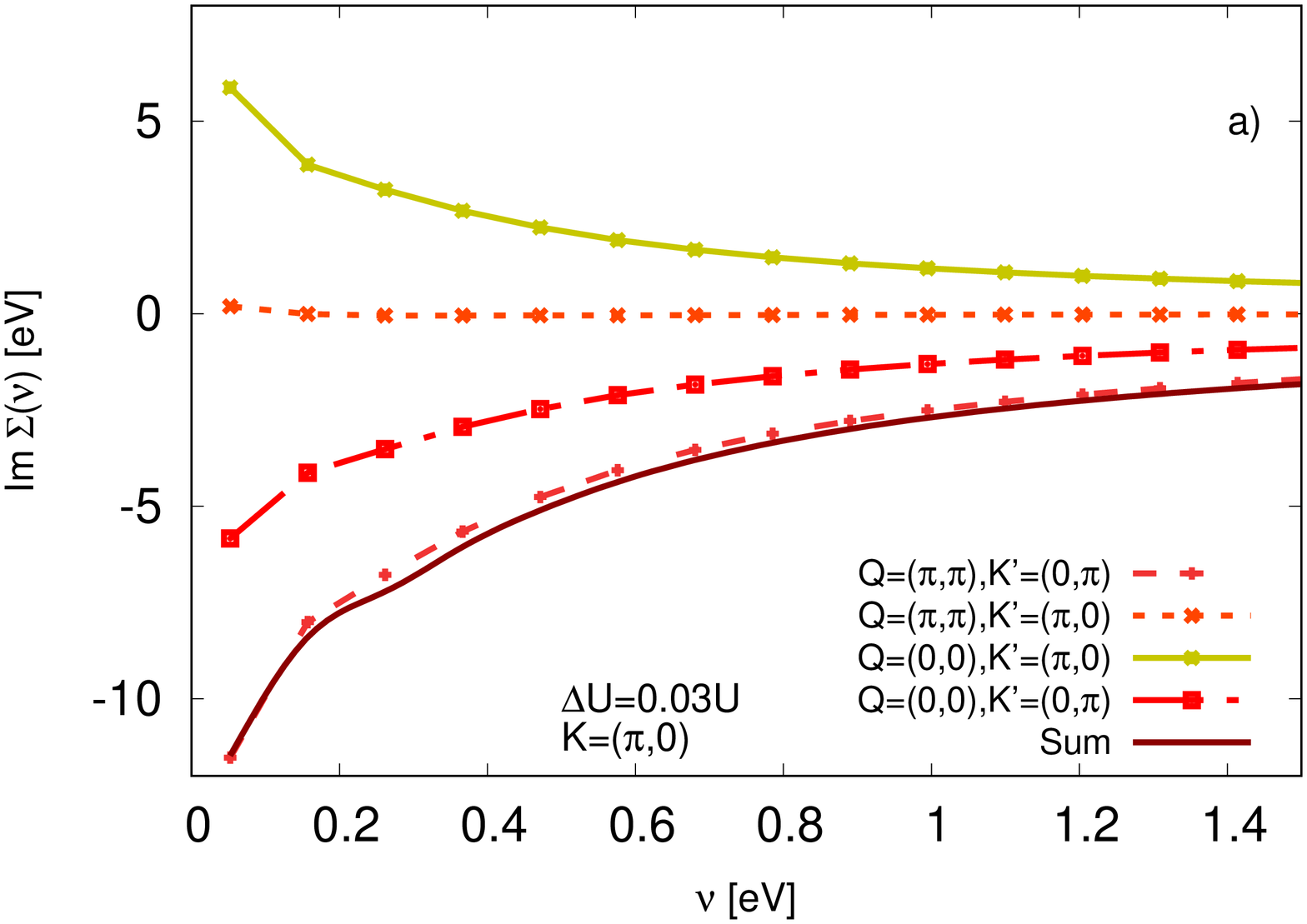}}}}
{\rotatebox{-90}{\resizebox{6.0cm}{!}{\includegraphics {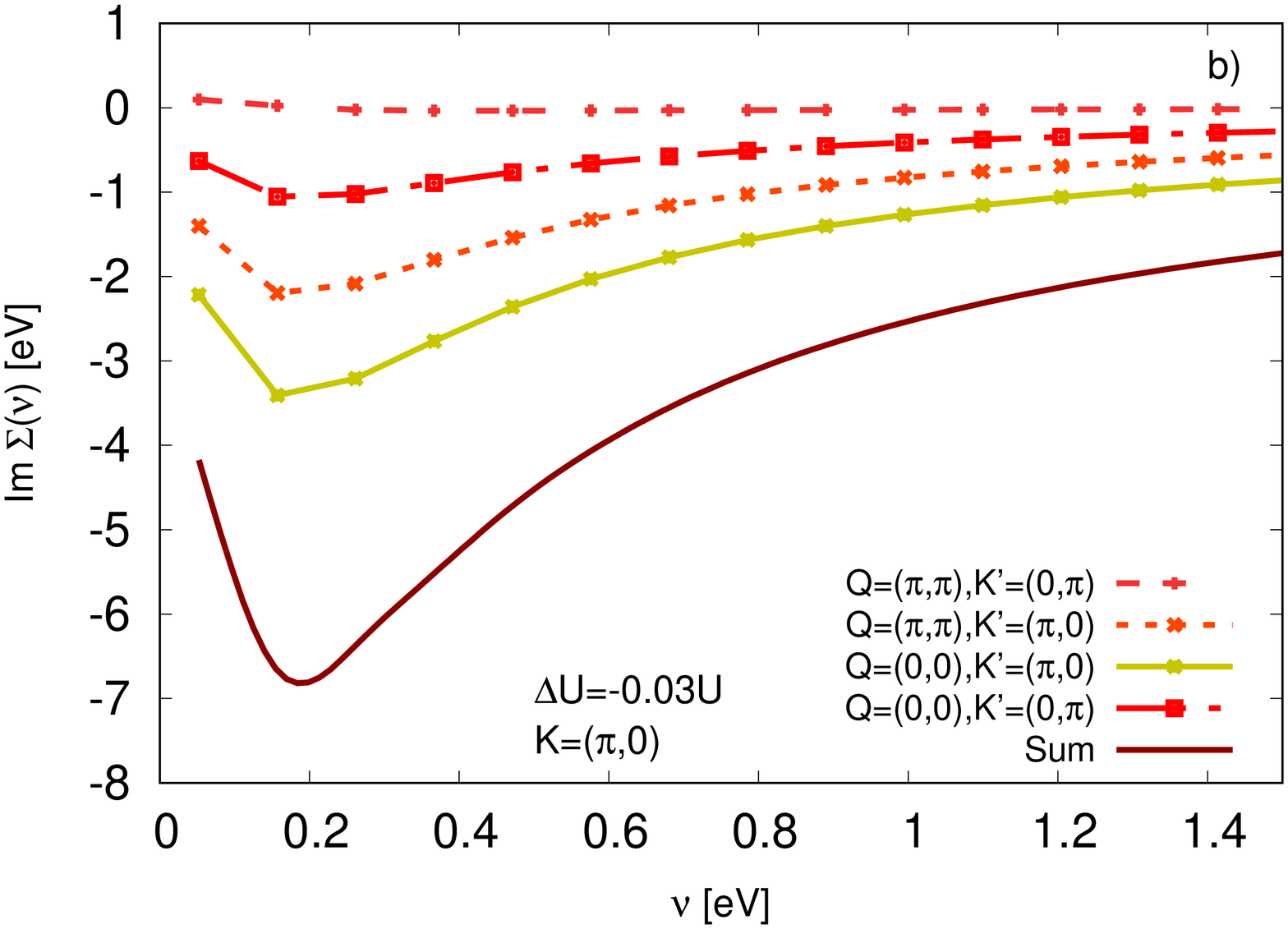}}}}
\caption{Contributions to the self-energy of the four-level model for ${\bf K}=(\pi,0)$, labeled 
according to Eqs.~(\ref{eq:11}, \ref{eq:11a}) as well as the sum ("Sum") 
of these contributions for $\Delta U=0.03$ U (a) and $\Delta U=-0.03$ (b). 
The figure illustrates the great importance of the sign of $\Delta U$ both 
for the shape of $\Sigma(\nu)$ and for the relative contributions of the 
different terms Eq.~(\ref{eq:dse_a}). The term for ${\bf Q}=(\pi,\pi)$ and 
${\bf K}'=(0,\pi)$ gives the main contribution to $\Sigma$ for $\Delta U>0$, 
but is almost zero for $\Delta U<0$.  The parameters are $V=-0.1$ eV, 
$U=1.6$ eV and $\beta=60$ eV$^{-1}$.
}\label{fig:2}
\end{figure}

We first consider the four-level model for $\Delta U>0$. Fig.~\ref{fig:1} 
shows various correlation functions  for ${\bf K}=(\pi,0)$ (also referred 
to as level $1c$). For small values of $U$ these are all very small. In 
the case of $A$, $B$, $C_{\rm sp}$ and $D$ this is partly because of a 
prefactor $U$ in Eq.~(\ref{eq:11a}), but these quantities remain small 
even if the prefactor $U$ is divided out. In fact, in the $U
\rightarrow 0$ limit 
the interaction between the two ${\bf K}$-orbitals on the cluster
vanishes, 
and each ${\bf K}$-orbital only interacts with its bath. Therefore these
correlation functions would be zero even if the factor $U$ were divided out. 

As $U$ is increased, there is also interaction between the cluster  
${\bf K}$-orbitals, but the interaction with the bath initially 
dominates. The increase of $U$ at first leads to a reduction of 
the double occupancy of each ${\bf K}$, as indicated by $L[(\pi,0)]$ 
[Eq.~(\ref{eq:3.1a})] becoming negative. This means that a spin 1/2 
state starts to develop in each ${\bf K}$-orbital. This spin is found to have 
an antiferromagnetic correlation to a spin in its bath.\cite{Jaime} Thus a 
Kondo-like state starts to develop for ${\bf K}=(\pi,0)$ and for ${\bf K}=
(0,\pi)$. The corresponding spectrum has a peak at the Fermi energy.\cite{Jaime}

For intermediate values of $U\sim 1$ eV, $L[(\pi,0)]$ turns positive,
reflecting the fact that the double occupancy of each ${\bf K}$-orbital is now enhanced,
instead of being suppressed. The reason is that a state of the type in 
Eqs.~(\ref{eq:5.5}, \ref{eq:5.2a}) starts to develop, for which each 
${\bf K}$-orbital has an enhanced double occupancy. This is an RVB-like
state [Eq.~(\ref{eq:72})], which is strongly correlated. In fact, as $U$ is
increased the energy gain from the Kondo effect is reduced and at the same 
time the energy gain from the formation of the RVB-like state in Eqs.~(\ref{eq:5.5}, 
\ref{eq:5.2a}) increases, tipping the balance to the RVB-like state.\cite{Jaime} 

We proceed  now by  analyzing  the corresponding spectral function, first in the language of Eq.~(\ref{eq:1.2}), following
Ref.~\onlinecite{Jaime}. For $\Delta U>0$, if an electron is removed from the 
cluster in a photoemission process, the RVB-like state is broken up. An electron can 
then hop in from the bath and form an RVB-like state again. In a similar way, if an electron 
is added in an inverse photoemission process, an extra electron can hop to the bath, 
leaving an RVB-like state on the cluster behind. The result is then spectral intensity 
at the Fermi energy. The crucial question is then what is the probability for 
forming RVB-like states in the final states after a photoemission or an inverse 
photoemission process.  There are two different paths coupling 
the ground-state to the final states corresponding to intensity 
close to the Fermi energy:\cite{Jaime} (i) For a localized non-degenerate state on the cluster 
($\Delta U>0$), like Eqs.~(\ref{eq:5.5}, \ref{eq:5.2a}), there is destructive interference 
and a pseudogap at the Fermi energy; (ii) For a localized degenerate state ($\Delta U<0$), 
the interference is constructive, leading to to a peak at the Fermi 
energy.\cite{Ferrero07,Leo04,Capone,Jaime}

We next discuss the consequences of the formation of the state in 
Eqs.~(\ref{eq:5.5}, \ref{eq:5.2a}) in terms of the self-energy and 
the language of Eq.~(\ref{eq:1.1a}). As discussed in Sec.~\ref{sec:4.1} 
below Eq.~(\ref{eq:72}), the formation of such a state leads to large 
negative values for the correlation function $A_{\uparrow \downarrow}[{\bf K},
{\bf K}+(\pi,\pi),(\pi,\pi)]$. This in turn leads to a large negative contribution 
to $B_{\uparrow \downarrow}[{\bf K},{\bf K}+(\pi,\pi)]$, as it  would be 
expected for an RVB-like state according to the discussion in Sec.~\ref{sec:3.3}. 
Finally, this is also responsible for the main contribution to $D_{\uparrow \downarrow}
({\bf K})$, which approaches the value -1 for large $U$. All together,
this leads to a 
large negative value for Im $\Sigma({\bf K},\nu_n)$ for small $\nu_n$ 
[Eq.~(\ref{eq:2.10c})] and, hence, to a small value for the spectral function at the Fermi energy. 

Our numerical calculations of the correlation functions $A$, $B$ and $D$  are reported in Fig.~\ref{fig:1}, as a 
function of increasing $U$, while results for the self-energy $\Sigma$ of the four-level model are shown 
in Fig.~\ref{fig:2}a for a large value of $U$ and  $\Delta U>0$. Im $\Sigma$  is indeed very negative for small $\nu_n$. The figure illustrates that 
the main contribution comes from $A_{\uparrow \downarrow}
[(\pi,0),(0,\pi),(\pi,\pi)]$, as could also be seen in Fig.~\ref{fig:1}.
The results in Fig.~\ref{fig:2}a can be compared with Table~\ref{table:2},
where $A[(\pi,0),(0,\pi),(\pi,\pi)]$ dominates. Table~\ref{table:2} also 
shows two smaller contributions for ${\bf Q}=(0,0)$ with opposite signs. 
The corresponding two curves in Fig.~\ref{fig:2} also have similar magnitude 
with opposite signs. 

Fig.~\ref{fig:2}b shows results for $\Delta U<0$. These results are completely different 
from the results in Fig.~\ref{fig:2}a, but agree with Table~\ref{table:2a}.
In this case the (absolute) largest contribution comes from $A[(\pi,0),(\pi,0),(0,0)]$. 
The contribution from $A[(\pi,0),(0,\pi),(\pi,\pi)]$, which dominates for
$\Delta U>0$, is here very small. In this case, $\Sigma$ behaves as in a Fermi liquid. 
This large difference can be traced back\cite{Jaime} to the degenerate character of the 
lowest cluster state for $\Delta U<0$, in contrast to the non-degenerate character for $\Delta U>0$.

These two (opposite) cases illustrate how the spectrum can depend very sensitively on small 
modifications in the physical parameters, provided that such modifications lead to a change of ground-state. 
In Appendix~\ref{sec:A1} we give detailed formulas for the spectra. 

\subsection{$N_c=4$}\label{sec:5.2}

\begin{figure}
{\rotatebox{-90}{\resizebox{6.0cm}{!}{\includegraphics {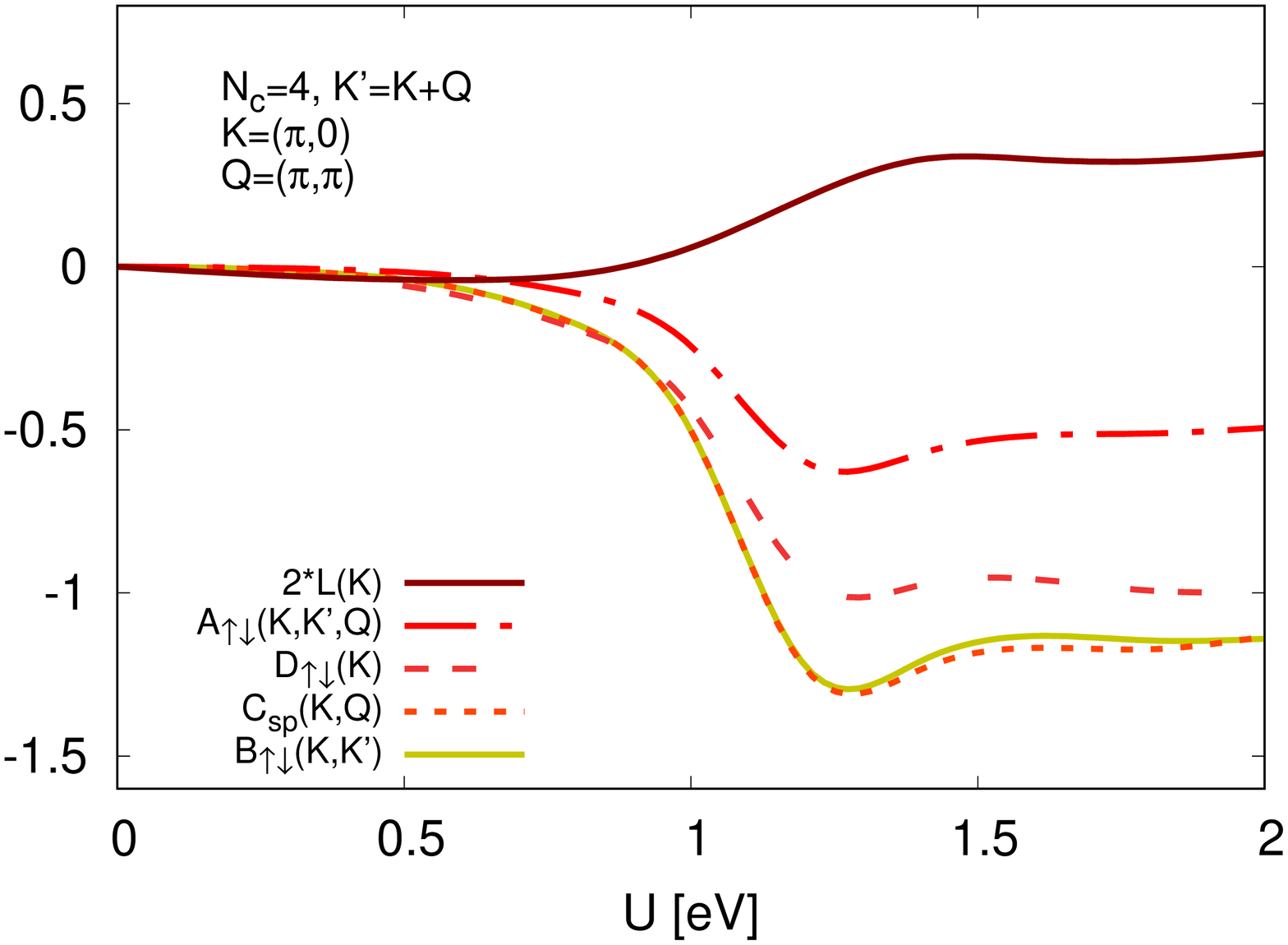}}}}
\caption{
Real parts of $B_{\uparrow \downarrow}({\bf K},{\bf K}')$, $C_{\rm sp}({\bf K},{\bf Q})$, and 
$D_{\uparrow \downarrow}({\bf K})$ for ${\bf K}=(\pi,0)$ and for ${\bf K}'=(0,\pi)$. 
The figure also shows Re $A_{\uparrow \downarrow}({\bf K},{\bf K}',{\bf Q})$ for ${\bf Q}=(\pi,\pi)$ as well as $L({\bf K})$. 
The parameters of the corresponding DCA calculation are $N_c=4$,
$t=-0.25$ eV and $\beta=60$ eV$^{-1}$. Note that the minor difference
between the equivalent values of $B_{\uparrow \downarrow}$ and
$C_{\rm sp}$ visible for $U> 1$ is a consequence of the finite
numerical precision of the calculation.
}\label{fig:3}
\end{figure}

\begin{figure}
{\rotatebox{-90}{\resizebox{6.0cm}{!}{\includegraphics {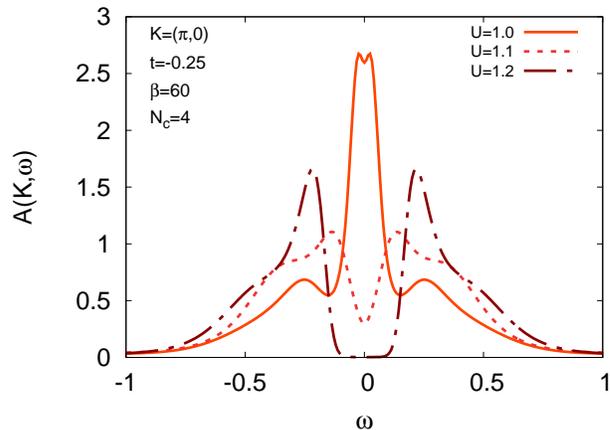}}}}
\caption{DCA Spectra for $U=1.0$, 1.1 and 1.2 eV.
The parameters are $N_c=4$, $t=-0.25$ eV and $\beta=60$ eV$^{-1}$.
}\label{fig:4}
\end{figure}

We now turn to the embedded clusters of DCA and study the case $N_c=4$ first.
Fig.~\ref{fig:3} shows the quantity $B_{\uparrow \downarrow}[(\pi,0),(0,\pi)]$, describing the 
coupling to an RVB-like state, $C_{\rm sp}[(\pi,0),(\pi,\pi)]$, used in the fluctuation 
diagnostics, and $D_{\uparrow \downarrow}[(\pi,0)]$, describing the total contribution to 
$\Sigma$, its largest component $A_{\uparrow \downarrow}[(\pi,0),(0,\pi),(\pi,\pi)]$ and the 
correlation function $L[(\pi,0)]$. These are the same correlation functions  
as shown in Fig.~\ref{fig:1} for the four-level model. The behavior is very
similar. This is not surprising. For large values of $U$ a localized state of
the approximate form in Eq.~(\ref{eq:5.2}) is formed,\cite{Jaime,Parquetprb}
which is identical to the localized state in Eq.~(\ref{eq:5.2a}) for the four-level
model, except for the occupied ${\bf K}=(0,0)$ levels. In fact, by
a systematic study\cite{Jaime}  of the correlation functions, it was found that the system also has a similar crossover from
a Kondo-like system to a localized system, as for the four-level model.

The largest changes in the correlation
functions considered (see Fig.~\ref{fig:3}) take place between $U=1.0$ eV and $1.2$ eV.
Fig.~\ref{fig:4} shows the corresponding spectra for $N_c=4$. For $U=1.0$ eV there is a peak at 
the Fermi energy, although there are signs of a pseudogap starting to develop,
while for $U=1.1$ eV there is a pseudogap and for $U=1.2$ eV a gap. This illustrates 
that the pseudogap opens up for the values of $U$ where $|B_{\uparrow 
\downarrow}|$ (and, thus,  $|C_{\rm sp}|$) become large. For $U=1.2$ eV,
$D[(\pi,0)]\approx -1$, consistent with a very large negative imaginary part of 
$\Sigma$ and the opening up of a gap, as discussed below  Eq.~(\ref{eq:2.10c}).  
For large $U$, the main contribution to $D_{\uparrow \downarrow}[(\pi,0)]$ comes 
from $B_{\uparrow \downarrow}[(\pi,0),(0,\pi)]$, and the contributions from 
$B_{\uparrow \downarrow}[(\pi,0),{\bf K}']$ for ${\bf K}'\ne {\bf K}+(\pi,\pi)$
are small. Similarly, if $D$ is expressed as a transfer momentum sum of $C_{\rm sp}$, the main
contribution to  $D[(\pi,0)]$ comes from $C_{\rm sp}[(\pi,0),(\pi,\pi)]$ 
and the contributions from $C_{\rm sp}[(\pi,0),{\bf Q}]$ for ${\bf Q}\ne (\pi,\pi)$
are small. 

Just as we could relate the properties of the four-level model 
to the formation of the localized state in Eq.~(\ref{eq:5.2a}),
the crucial common denominator for $N_c=4$ is the formation of the state in  
Eq.~(\ref{eq:5.2}). This leads to a positive value of $L[(\pi,0)]$ 
and shows how $|A_{\uparrow \downarrow}[(\pi,0),(0,\pi), (\pi,\pi)]|$ gets 
large at the same time as $L[(\pi,0)]$ goes positive, for reasons already 
discussed for the four-level model. Thus this specific component of $A$
provides a substantial contribution to $B_{\uparrow \downarrow}[(\pi,0), (0,\pi)]$, being 
about half of $B$. That $|A_{\uparrow \downarrow}[(\pi,0), (0,\pi),(\pi,\pi)]|$ 
becomes large when the state Eq.~(\ref{eq:5.2}) is formed can also be deduced from
the definition in Eq.~(\ref{eq:11a}). $|A_{\uparrow \downarrow}[(\pi,0),(0,\pi), 
(\pi,\pi)|$ also gives a substantial contribution $C_{\rm sp}[(\pi,0),{\bf Q}]$. 
At the same time, that $C_{\rm sp}$ must be also large can be inferred from Eq.~(\ref{eq:19g}), relating 
$C_{\rm sp}$ to antiferromagnetic spin correlations and from the fact that 
the RVB-like state in Eq.~(\ref{eq:5.2}) implies, per definition, an appreciable amount of 
this specific correlation.\cite{Jaime}

On the basis of these considerations, it is clear that the formation of the
state in Eq.~(\ref{eq:5.2}) controls, simultaneously, several correlation functions: (i) $L$, used in
Ref.~\onlinecite{Jaime}, (ii) $C_{\rm sp}$, introduced in Ref.
\onlinecite{Fluct}, as well as (iii) the related quantity $B$,  introduced  here to
describe more explicitly the coupling to RVB-like real space correlations.

Depending on the different theoretical perspectives, the origin
of the pseudogap observed in the DCA spectra of the $2d$ Hubbard model
can be ascribed, complementarily, either to antiferromagnetic spin correlations
or RVB-like correlations. 
We must stress here, however, that this is not a contradiction, since the RVB
state, as discussed above, does
lead to  substantial antiferromagnetic spin correlations, though
weaker than in a pure N{\'e}el state (for instance, for a cluster with $N_c=8$ and a large $U$ the difference in strength is about a factor of 
two.\cite{JaimearX}) Thus, in this regime, the two pictures (AF or
RVB correlations) simply provide different perspectives on the same physics.

We notice that $A_{\uparrow \downarrow}[{\bf K},{\bf K}+(\pi,\pi),(\pi,\pi)]$
gives almost the full contribution to $D$ in the four-level model but only
about half the contribution in the $N_c=4$ model (see Fig.~\ref{fig:3}),
the other 15 combinations of ${\bf K}'$ and ${\bf Q}$ together contributing
the other half. This illustrates the need to switch to summed quantities,
such as $B$ and $C$, for larger values of $N_c$, as the individual contributions
$A$ are reduced with $N_c$. As discussed in Sec.~\ref{sec:3} these summed
quantities can be related to real space correlation functions, and they
therefore have a weaker dependence on $N_c$. The quantity $L$ used in
Ref.~\onlinecite{Jaime} suffers from similar problems as an individual
$A$, namely of becoming smaller as $N_c$ increases.

\subsection{$N_c=8$}\label{sec:5.3}

\begin{figure}
{\rotatebox{-90}{\resizebox{6.0cm}{!}{\includegraphics {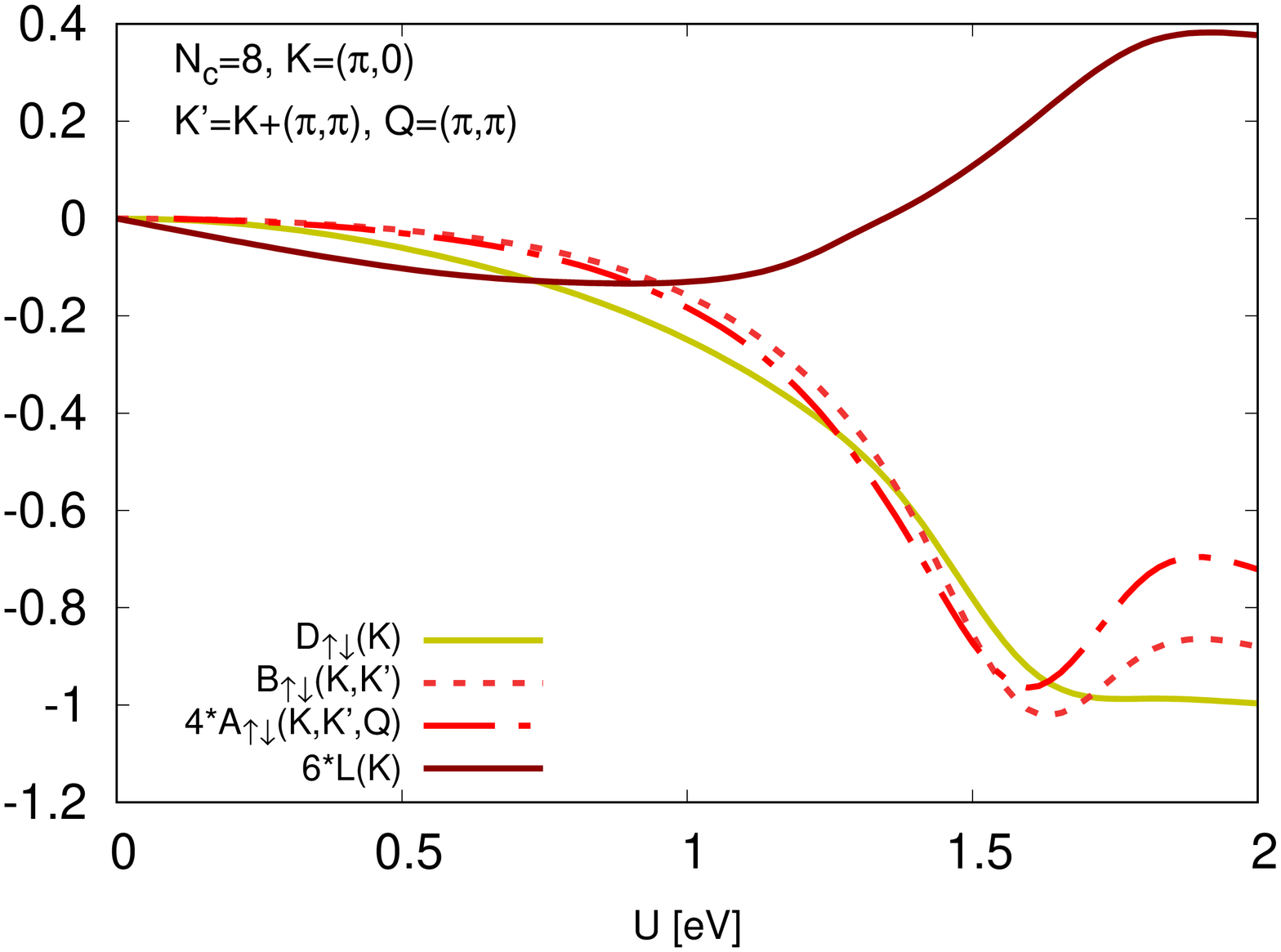}}}}
{\rotatebox{-90}{\resizebox{6.0cm}{!}{\includegraphics {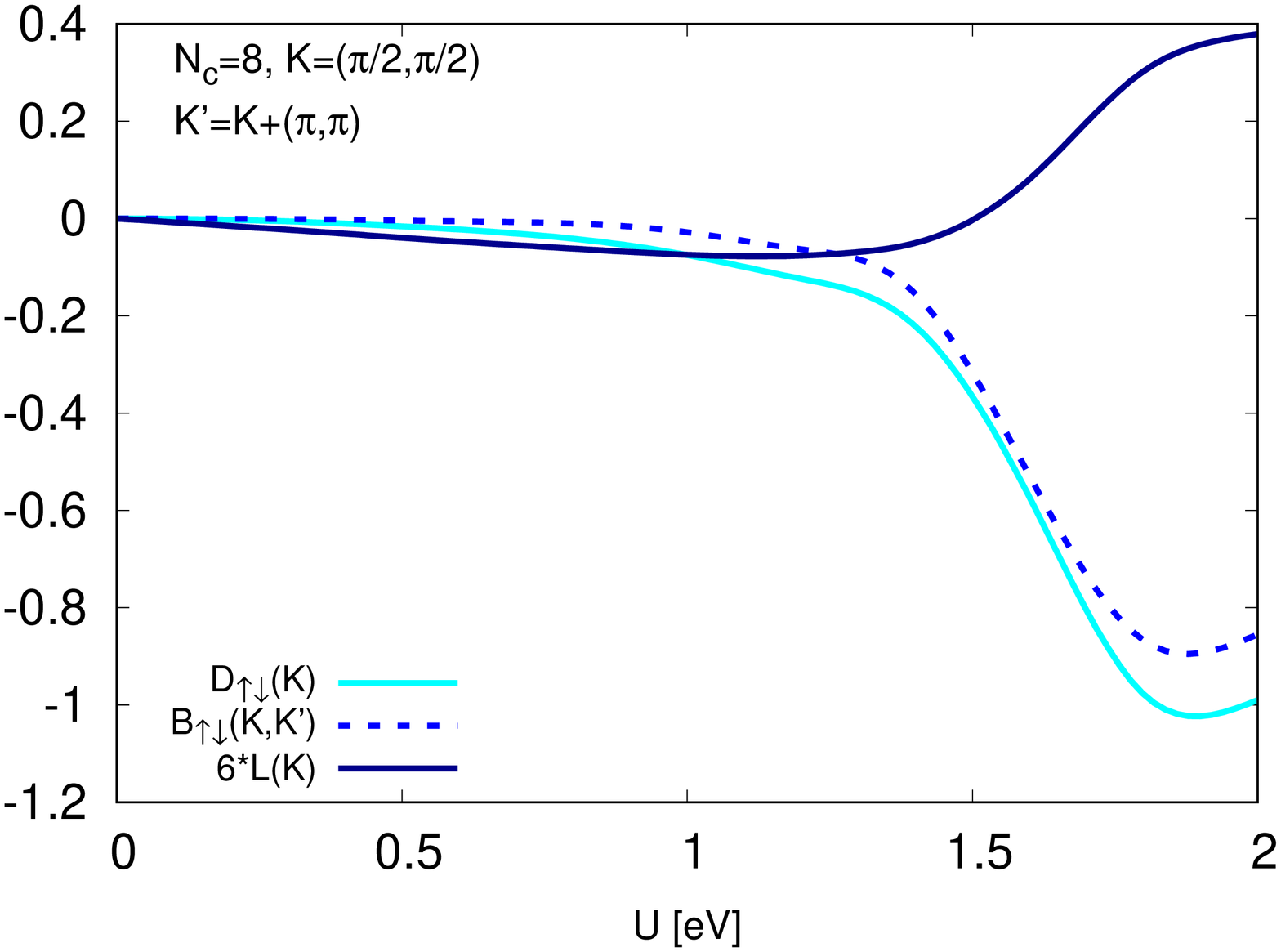}}}}
\caption{Re $B_{\uparrow \downarrow}({\bf K},{\bf K}')$ and 
Re $D_{\uparrow \downarrow}({\bf K})$ for ${\bf K}=(\pi,0)$ (upper panel, red, colors online) 
and $(\pi/2,\pi/2)$ (lower panel, blue colors online) and for ${\bf K}'={\bf K}+(\pi,\pi)$.
The parameters are $N_c=8$, $t=-0.25$ eV and $\beta=60$ eV$^{-1}$.
}\label{fig:5}
\end{figure}
\begin{figure}
{\rotatebox{-90}{\resizebox{6.0cm}{!}{\includegraphics {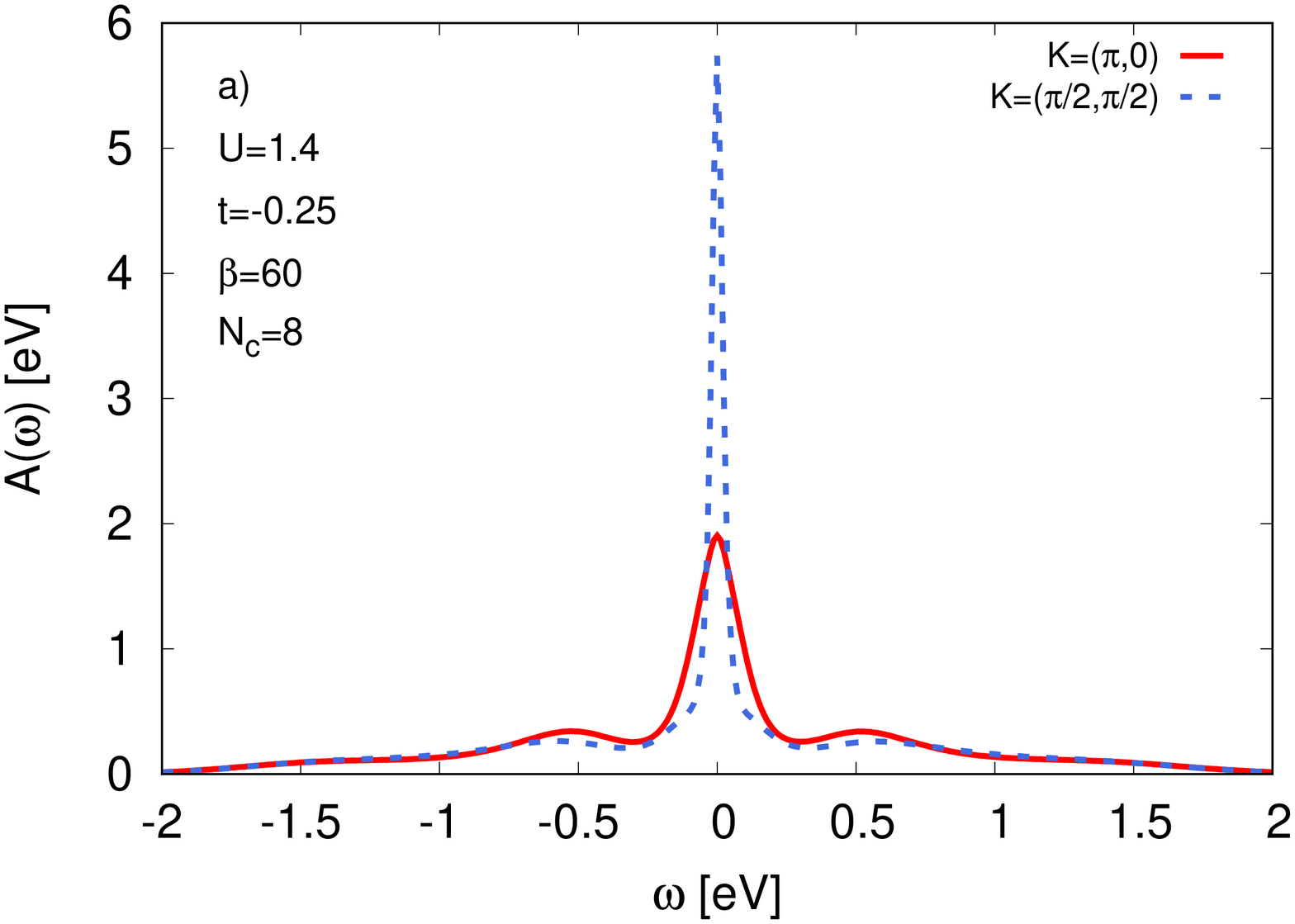}}}}
{\rotatebox{-90}{\resizebox{6.0cm}{!}{\includegraphics {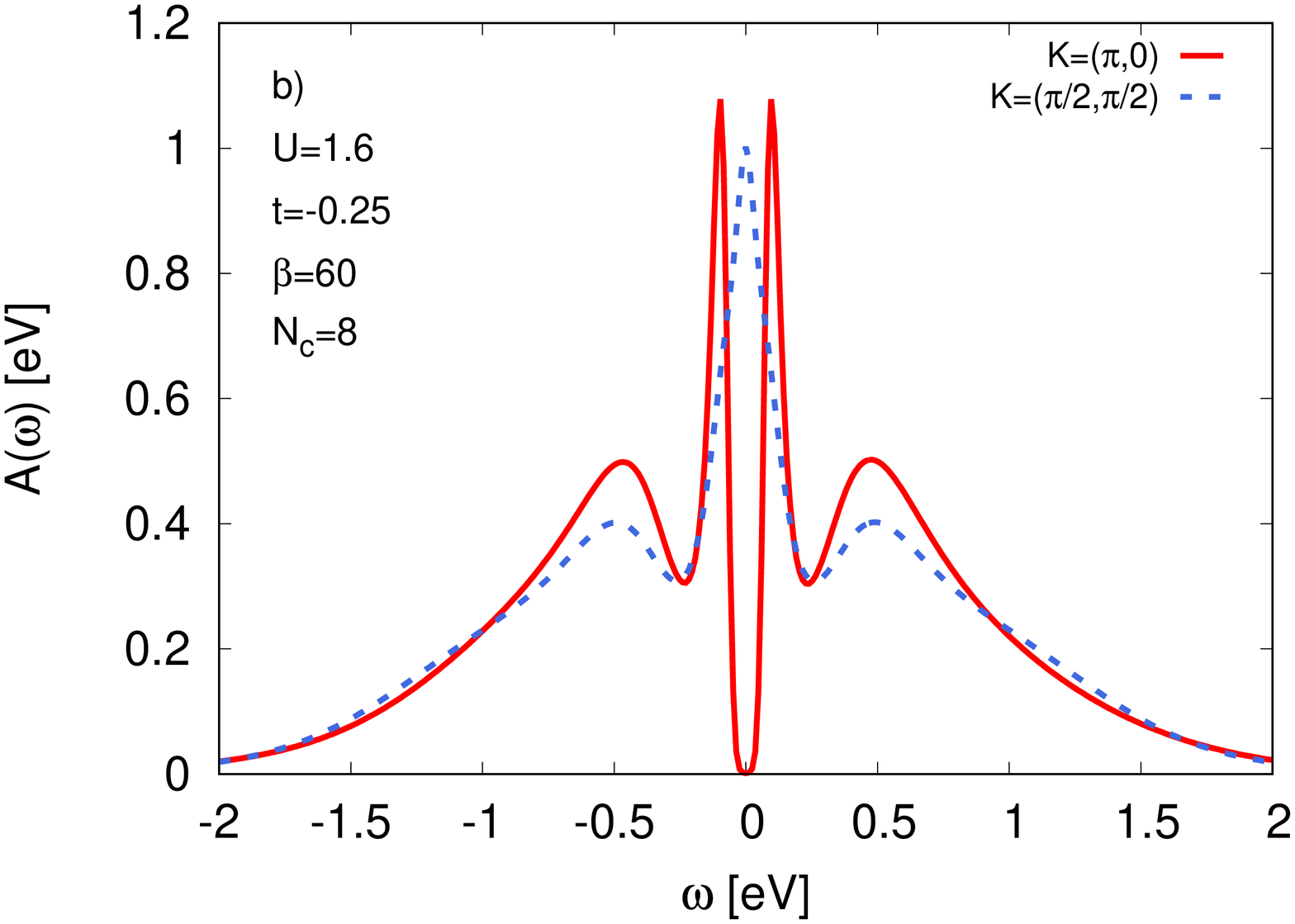}}}}
{\rotatebox{-90}{\resizebox{6.0cm}{!}{\includegraphics {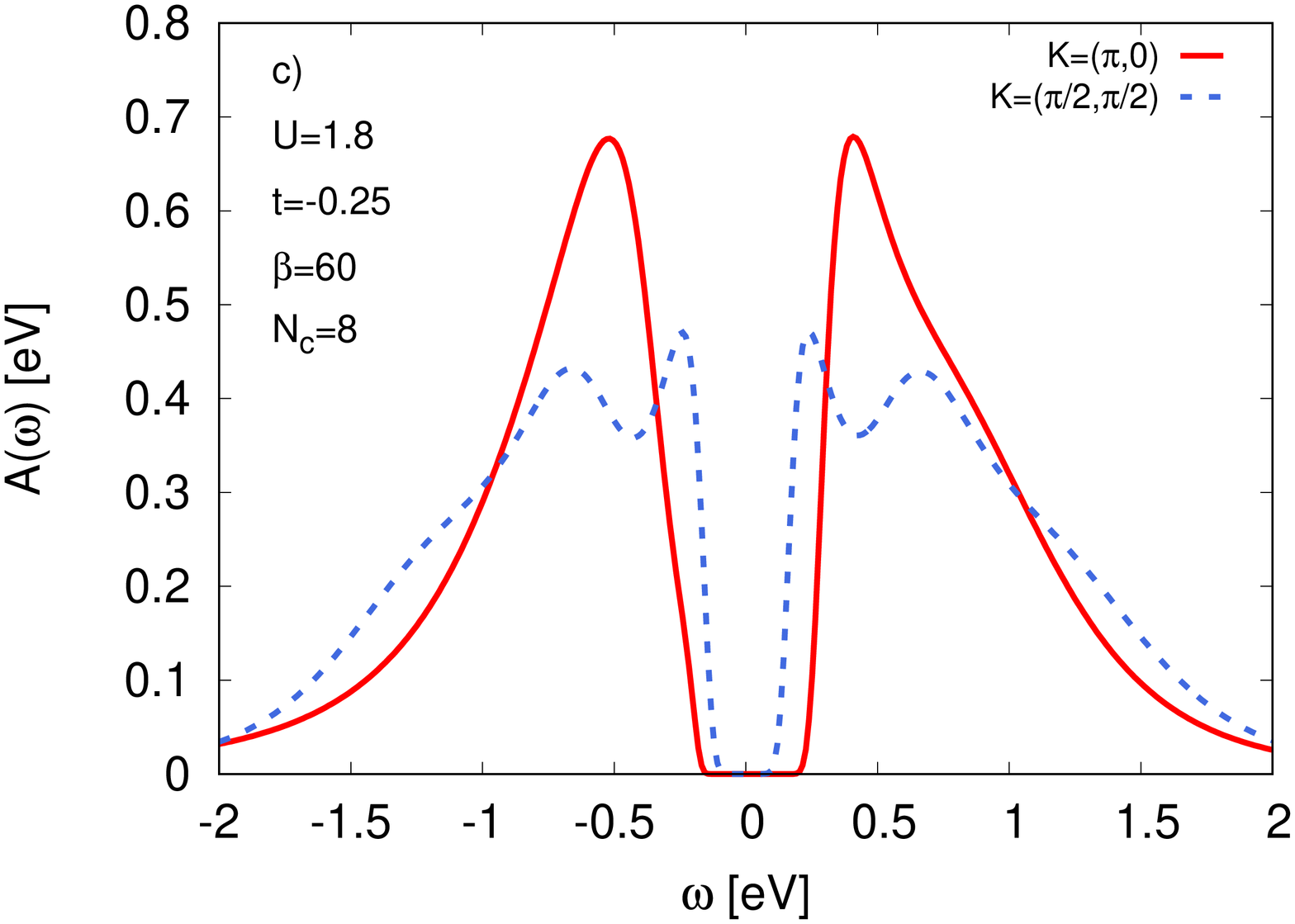}}}}
\caption{DCA Spectral functions as a function of ${\bf K}$ for $U=1.4$  (a),
1.6 (b)  and  $U=1.7$  (c). The parameters are $N_c=8$, $t=-0.25$ eV 
and $\beta=60$ (eV)$^{-1}$
\label{fig:6}}
\end{figure}

We now turn to $N_c=8$. Fig.~\ref{fig:5} shows correlation functions 
for ${\bf K}=(\pi,0)$ and $(\pi/2,\pi/2)$. Similar to the case $N_c=4$, $L$ 
first turns negative as $U$ is increased and then positive for larger values 
of $U$. As shown in Table~\ref{table:6.5} and emphasized in Ref.~\onlinecite{Jaime},
the coupling to the bath is much weaker for ${\bf K}=(\pi,0)$ and for $(0,\pi)$ than for
$(\mp \pi/2, \mp \pi/2)$. Therefore as $U$ is increased, first there is a switch for 
${\bf K }=(\pi,0)$ and $(0,\pi)$ from a Kondo-like state to a state where the $(\pi,0)$ 
and $(0,\pi)$ orbitals form a state similar to state (\ref{eq:5.2}). 
The loss of the Kondo energy in the ${\bf K}=(\pi,0)$ and $(0,\pi)$ channels is 
then more than compensated by the substantial energy gain from the correlation 
of the ${\bf K}=(\pi,0)$ and $(0,\pi)$ cluster orbitals.
Over a certain range of $U$ values, the orbitals for ${\bf K} =(\pm \pi/2,
\pm \pi/2)$, however, still form Kondo-like states with their baths,
and they are fairly uncorrelated with other ${\bf K}$-values.\cite{Jaime}
This is also illustrated by $L[(\pi,0)]$ going from negative to positive values, 
while $L[(\pi/2,\pi/2)]$ still remains negative over some range of $U$ values.
As a result one observes\cite{Jaime} a pseudogap for ${\bf K}=(\pi,0)$
but not for ${\bf K} =(\pm \pi/2,\pm \pi/2)$. For larger values of $U$ also the 
Kondo states for ${\bf K}=(\pm \pi/2,\pm \pi/2)$ are lost and all the cluster orbitals 
become strongly correlated, forming a localized state. At this point $L[\pi/2,\pi/2)]$ 
also turns positive. Then a spectral gap also opens up for ${\bf K}=(\pm \pi/2,\pm \pi/2)$. 

Fig.~\ref{fig:5} clarifies the progression of the
physics with increasing interaction. When $L[(\pi,0)]$ turns positive, $B_{\uparrow \downarrow}[(\pi,0),
(0,\pi)]$ becomes very negative. This is due to the coupling to a state similar
to the state of Eq.~(\ref{eq:5.2}).  The orbitals  $(\pm \pi/2,\pm\pi/2)$ are also occupied,
but they are more entangled with their baths rather than with the
cluster states  $(\pi,0)$ and $(0,\pi)$.\cite{Jaime}  At this point 
correlation functions involving $(\pi,0)$ and $(0,\pi)$ take values similar to the 
ones in the RVB state.  For somewhat larger values of $U$,  $L[(\pm \pi/2,\pm \pi/2)]$ 
turns positive.  Then $B_{\uparrow \downarrow}[(\pi/2,\pi/2),(-\pi/2,-\pi/2)]$ also
becomes very negative, and correlation functions involving Fermi surface momenta approximately take on their values for 
an RVB state. A proper RVB state, however, does not develop until $U$ becomes substantially larger, as discussed 
in Appendix \ref{sec:5.6}. 

Fig.~\ref{fig:6} shows the spectra for ${\bf K}=(\pi,0)$ and $(\pi/2,\pi/2)$
and for different values of $U$. In particular, the figure illustrates how a 
pseudogap appears  for ${\bf K}=(\pi,0)$ and $U=1.6$ eV. In Fig.~\ref{fig:5} 
$|B_{\uparrow \downarrow}[(\pi,0),(0,\pi)]|$ and $L[(\pi,0)]$ indeed become large for this value of $U$,
signaling a localization in the $(\pi,0)$-$(0,\pi)$ space which causes the
pseudogap. For $U=1.8$ eV, Fig.~\ref{fig:6}c shows how a spectral gap
is opened  also for ${\bf K}=(\pi/2,\pi/2)$. In Fig.~\ref{fig:5} 
$|B_{\uparrow \downarrow}[(\pi/2,\pi/2),(-\pi/2,-\pi/2)]|$ and $L[(\pi/2,\pi/2)]$ become 
large for this value of $U$, signaling that a corresponding
localization  has also occurred. The evolution of spectra and correlations functions 
illustrates that the spectra are consistent with the arguments above.

It is important to note, however, that the picture emerging from
Fig.~\ref{fig:5} is different from the one where the system is viewed as a set of ${\bf K}$-orbitals coupling to their
baths but not to each other. In such a picture there would be a number of 
independent DMFT calculations for each ${\bf K}$ and a series of momentum selective    
Mott transitions. Correspondingly, one would see how the double occupancy 
of each ${\bf K}$-state is suppressed as $U$ is increased, and the gap forms 
because a gap develops also in the bath due to self-consistency. In Fig.~\ref{fig:5} 
the double occupancy in the ${\bf K}$-orbitals is also suppressed ($L({\bf K})<0$)for small $U$, 
but for larger $U$ the double occupancy is increased ($L({\bf K})>0$). This is due to the switch 
over from Kondo like states to a localized state with strong correlation between 
${\bf K}=(\pi,0)$ and $(0,\pi)$, as illustrated by the behavior of $L$, $A$ 
and $B$. This demonstrates that the correlation between different ${\bf K}$-states is 
crucial for the pseudogap formation. The formation of a pseudogap in
the bath happens also here,  due to the self-consistency of DCA. However, this is not     
a general prerequisite for developing a pseudogap in the spectrum of the embedded cluster. 
In fact, a pseudogap can also form,\cite{Jaime} if the bath, not calculated self-consistently,
is purely metallic, although a somewhat larger $U$ is now required. 
In this case, the pseudogap forms merely due to the strong correlation between $(\pi,0)$ and 
$(0,\pi)$. As $U$ increases, double occupancy of the {\it sites} is suppressed.
This {\it real} space correlation cannot be described in reciprocal space as a suppression of double 
occupancy of ${\bf K}$-states, but rather as a correlation between different ${\bf K}$'s.
It is then not surprising that this correlation becomes very important for large $U$.

\subsection{$N_c=32$}\label{sec:5.5}
\begin{figure}
{\rotatebox{-90}{\resizebox{6.0cm}{!}{\includegraphics {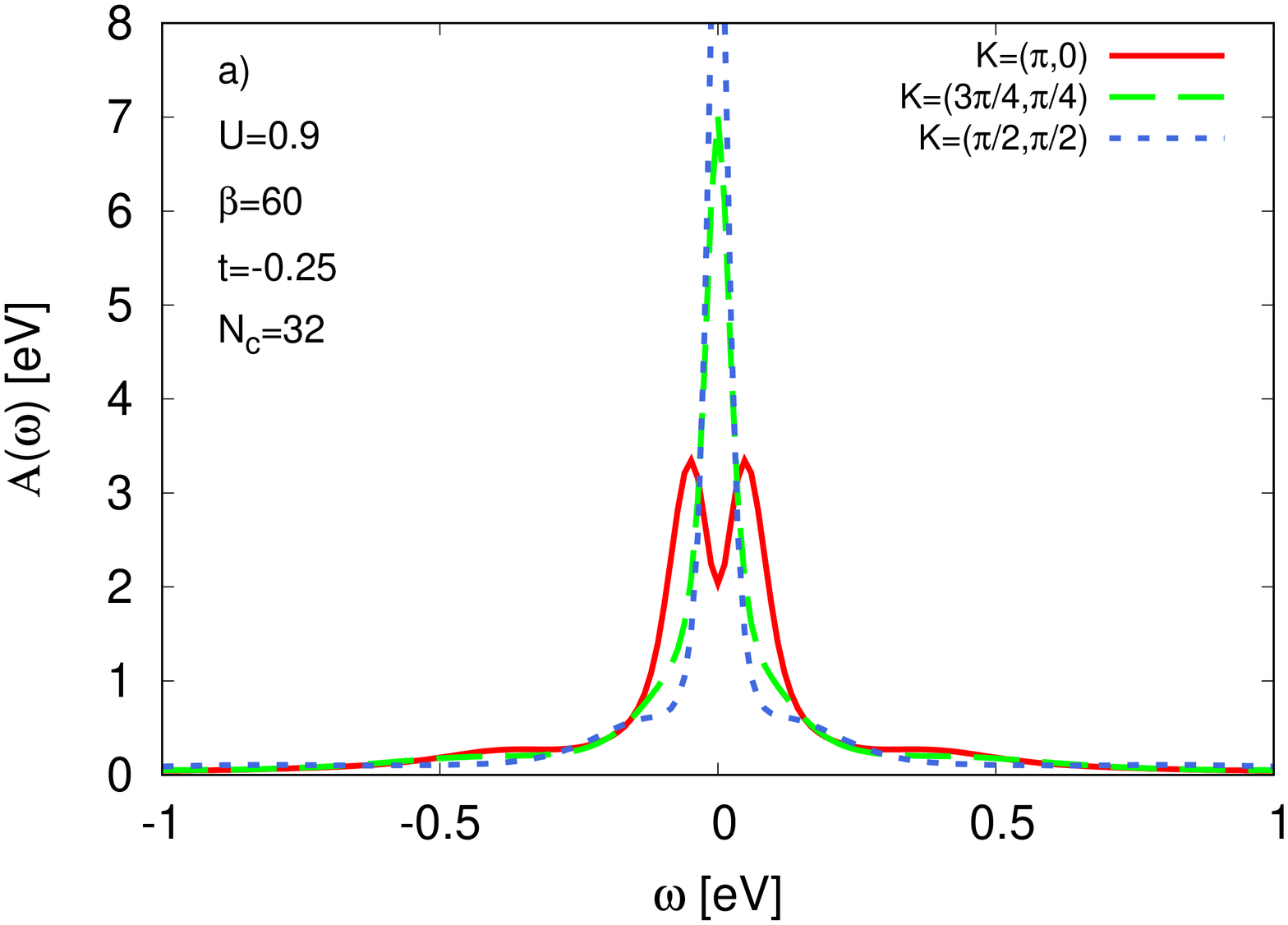}}}}\label{fig:7a}
\vskip-3.5cm
\hskip-3.7cm
{\rotatebox{-0}{\resizebox{2.0cm}{!}{\includegraphics {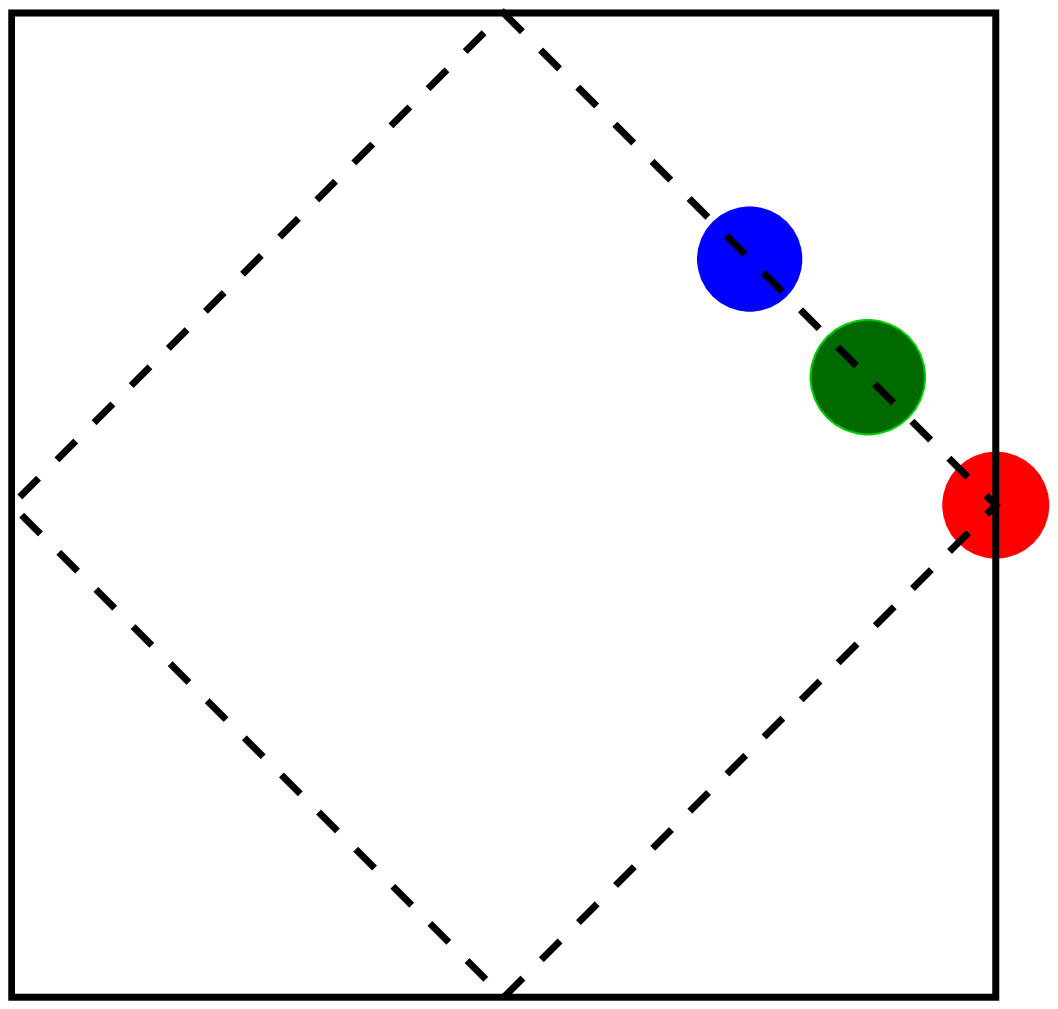}}}}
\vskip1.5cm
{\rotatebox{-90}{\resizebox{6.0cm}{!}{\includegraphics {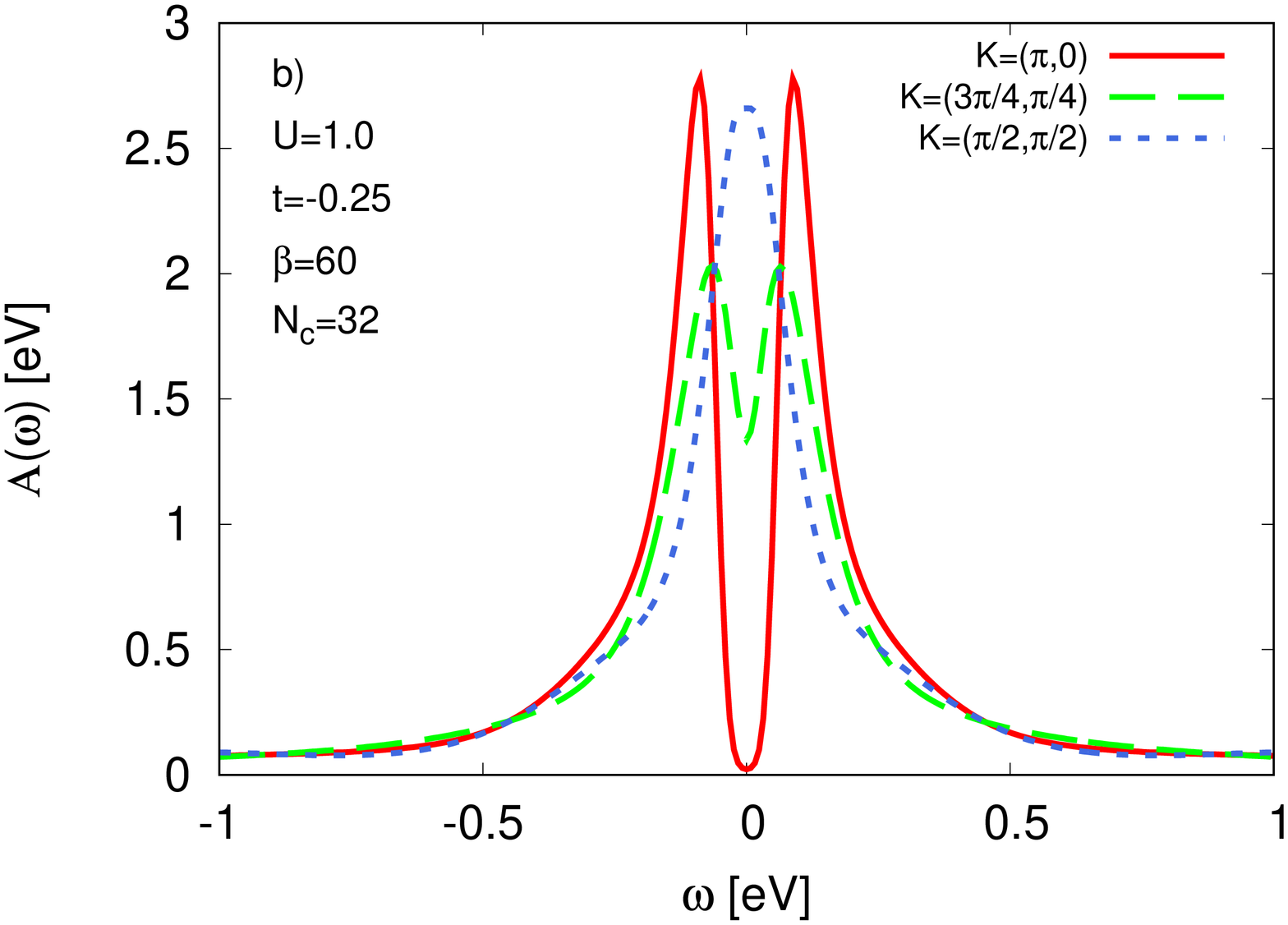}}}}
{\rotatebox{-90}{\resizebox{6.0cm}{!}{\includegraphics {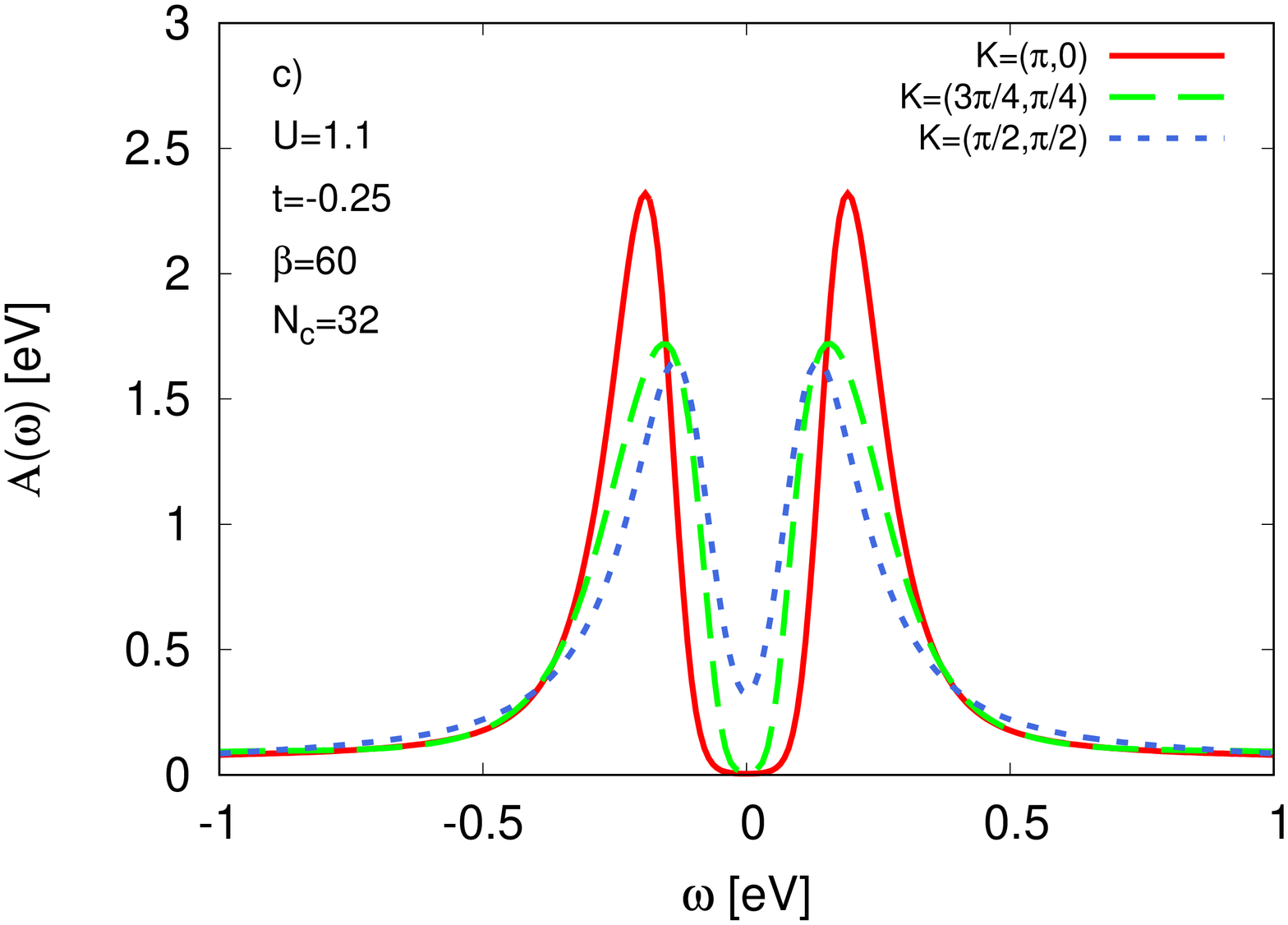}}}}
\caption{\label{fig:7}Spectral functions as a function of ${\bf K}$ for $U=0.9$ eV (a), $U=1.0$ eV (b) and $U=1.1$ eV (c). The parameters are $N_c=32$, $t=-0.25$ eV and $\beta=60$ (eV)$^{-1}$.
}
\end{figure}
\begin{figure*}
{\rotatebox{-90}{\resizebox{4.0cm}{!}{\includegraphics {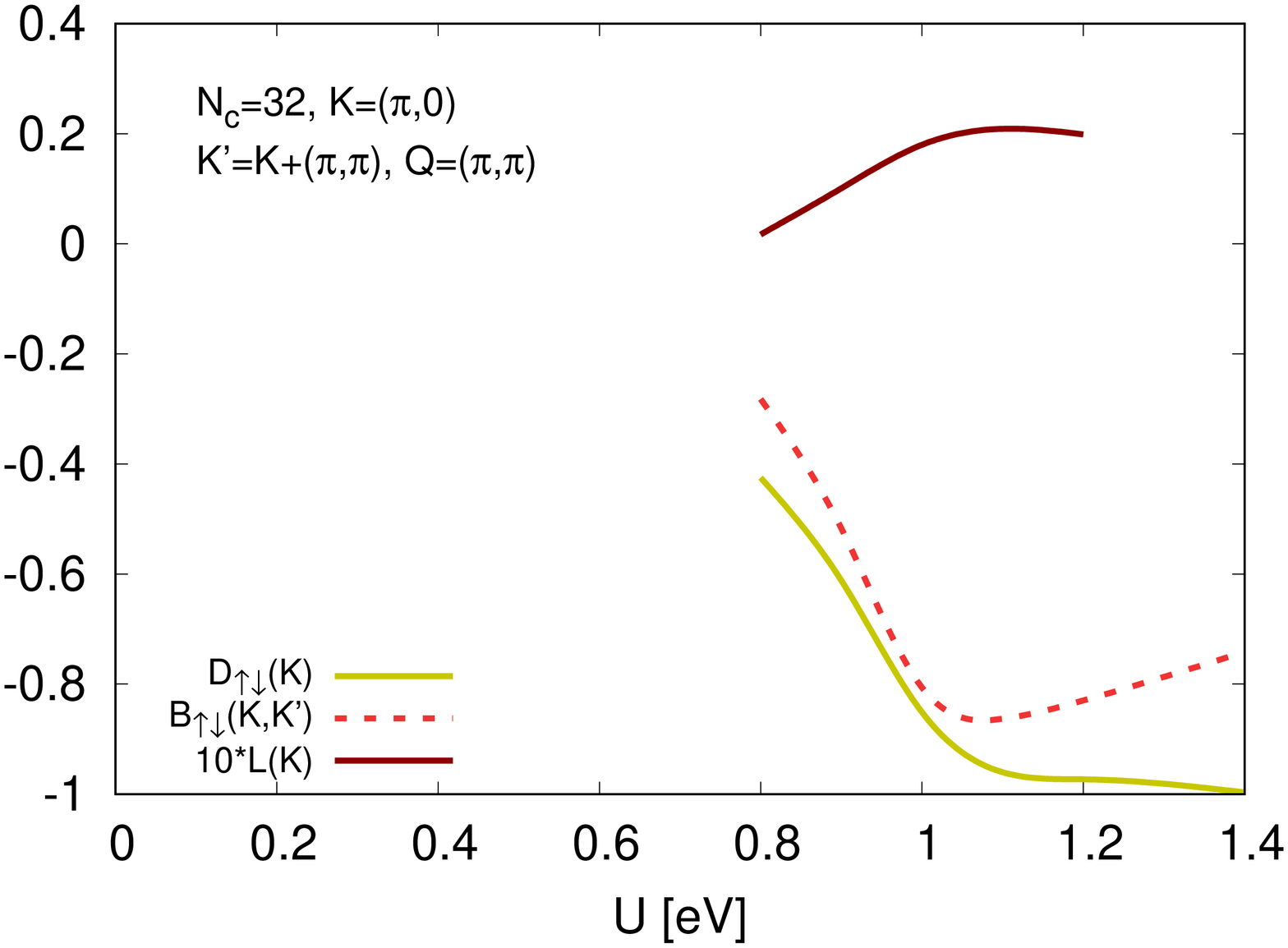}}}}
{\rotatebox{-90}{\resizebox{4.0cm}{!}{\includegraphics {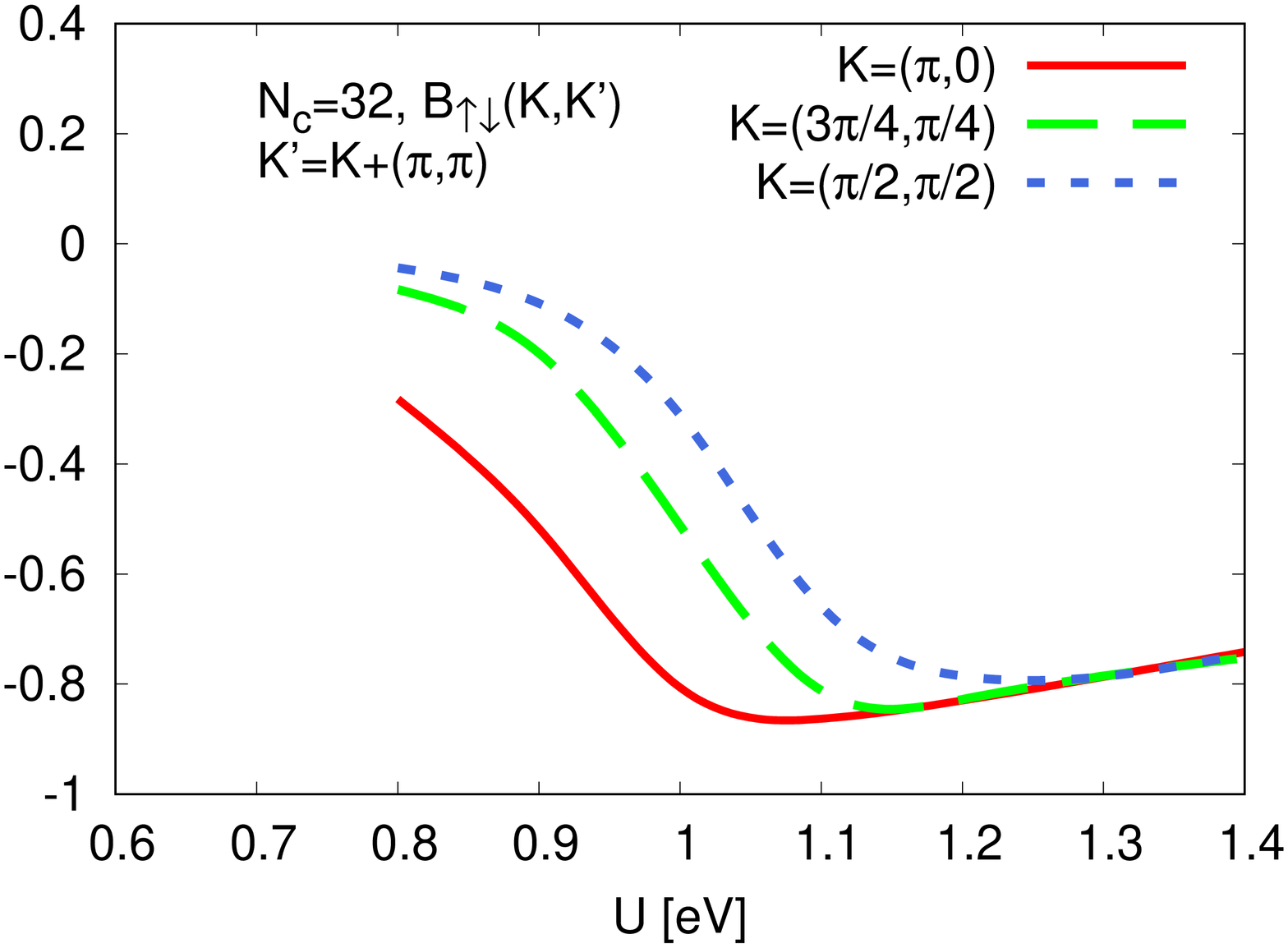}}}}
{\rotatebox{-90}{\resizebox{4.0cm}{!}{\includegraphics {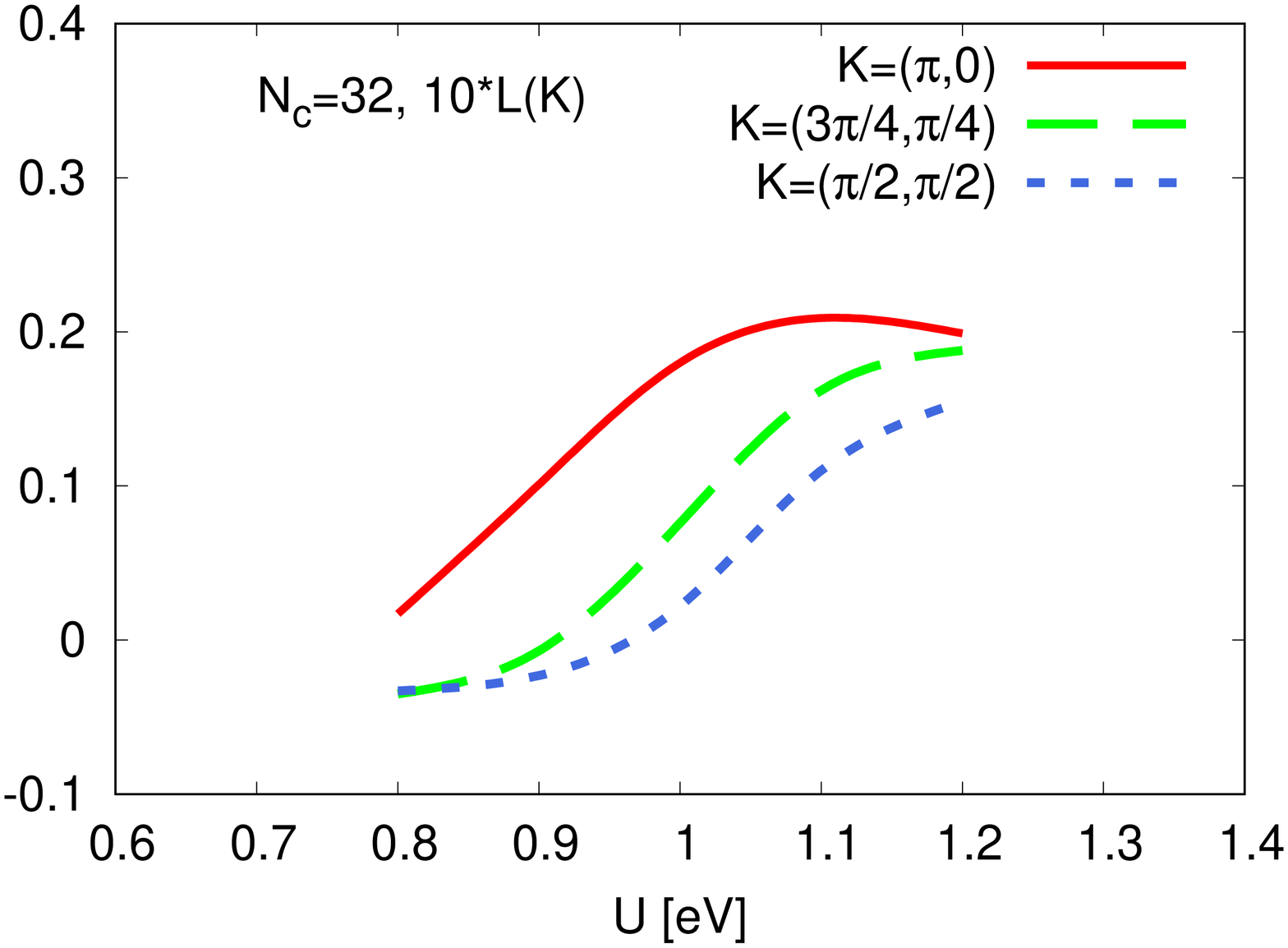}}}}
\caption{\label{fig:8} (colors online) 
Re $D({\bf K})$ (left panel), Re $B({\bf K},{\bf K}')$ (middle panel) and $L({\bf K})$ (right panel) for ${\bf K}=(\pi,0)$ (red solid lines), ${\bf K}=(3\pi/4,\pi/4)$ (green long-dashed lines) 
and ${\bf K}=(\pi/2,\pi/2)$ (blue short-dashed lines) and ${\bf
  K}'={\bf K}+(\pi,\pi)$. The parameters of the corresponding DCA
calculation are $N_c=32$, $t=-0.25$ eV and $\beta=60$ (eV)$^{-1}$.
}
\end{figure*}

While DCA calculation for larger clusters become numerically more
expensive, the additional effort is certainly worth it for the case of
a $N_c=32$ . In fact in this DCA cluster,  there are {\sl  three} inequivalent ${\bf K}$-points 
on the Fermi surface, $(\pi,0)$, $(3\pi/4,\pi/4)$ and $(\pi/2,\pi/2)$
(see inset of Fig.~\ref{fig:7}a), which makes it possible to study in more 
detail how the pseudogap evolves as $U$ is increased.

Fig.~\ref{fig:7} shows the spectra. For $U=0.9$ eV only ${\bf K}=(\pi,0)$ 
shows signs of a pseudogap, while there are still metallic peaks at $E_F$ for ${\bf K}=
(3\pi/4,\pi/4)$ and $(\pi/2,\pi/2)$. Increasing $U$ to 1.0 eV, there is 
also a pseudogap for ${\bf K}=(3\pi/4,\pi/4)$ and, eventually, for $U=1.1$ eV also 
${\bf K}=(\pi/2,\pi/2)$ shows an evident non Fermi-liquid behavior.

The observed progression can be understood by studying the couplings 
[Eq.~(\ref{eq:2.6})]  to the baths for these three ${\bf K}$ points,
as shown in Table~\ref{table:6.5}. As found before, the coupling 
for ${\bf K}=(\pi,0)$ is much weaker than for $(\pi/2,\pi/2)$. 
Table~\ref{table:6.5} shows that the point in between, ${\bf K}=
(3\pi/4,\pi/4)$ indeed has an intermediate coupling. The switch 
from a Kondo-type of states to a localized state then happens 
successively for the three ${\bf K}$ points as $U$ is increased.

This is  illustrated in Fig.~\ref{fig:8} showing
$D_{\uparrow \downarrow}({\bf K})$, $B_{\uparrow \downarrow}[{\bf K},
{\bf K}+(\pi,\pi)]$ and $L({\bf K})$. As for smaller $N_c$, the spectrum 
for ${\bf K}$ obtains a pseudogap shortly after $L({\bf K})$ turns 
positive. At the same time  $B_{\uparrow \downarrow}[{\bf K}, 
{\bf K}+(\pi,\pi)]$ and $D_{\uparrow \downarrow}({\bf K})$ become 
strongly negative. 

As $U$ is increased, correlation functions involving 
${\bf K}=(\pi,0)$ and $(0,\pi)$ first approach values corresponding to an RVB state.
At a somewhat later point this also applies for ${\bf K}=(\pm 3\pi/4,\pm \pi/4)$ 
and $(\pm \pi/4,\pm 3\pi/4)$ and for yet larger values of $U$ for $(\pm \pi/2,\pm \pi/2)$.
Thus the gradual development of an RVB state leads the gradual development
of a pseudogap for ${\bf K}$-points along the Fermi surface.

\section{Conclusions}\label{sec:6}

In this paper, we have studied the relations between electron spectral functions and real space
correlation functions. We used the Schwinger-Dyson equation to establish
the connections between the electron self-energy and two-particle vertex functions 
$F(k,k',q)$, involving summations over $k'$ and $q$ to obtain $\Sigma(k)$.
In fact, while $F$ contains a wealth of information about the scattering 
of the interacting particles, it is a challenging task to disentangle all the
effects of these scattering processes on
$\Sigma$, due to the intrinsic complexity of $F$. 
In this work, we have shown that this goal can be achieved through
different, but physically equivalent,  paths: Either one performs a sum over $k'$ in the equation
of motion for $\Sigma$, making explicit  the contributions as a function of the transferred energy/momentum $q$
(``bosonic fluctuation diagnostics'') or one sums over $q$ and study the contributions to $\Sigma(k)$ as a function of 
$k'$ (``fermionic fluctuation diagnostics''). While the correlation
functions of the two formulations are related by exact expressions, at
least for the SU(2)-symmetric case, the latter allows for
a  more natural interpretation in terms of real space RVB
fluctuations. 

To improve our physical understanding, we exploited a large $U$ approximation,  
which allows us to establish more direct, semi-analytical relations between the 
spectral function and real space correlations. Comparison with numerical calculations 
for small clusters shows that this approximation is sufficiently 
accurate for values of $U$ where a spectral gap has developed over the 
whole Fermi surface in our DCA results for the two-dimensional Hubbard
model. In this way, we could relate the spectra to real space 
charge, spin, superconductivity and RVB correlations. The approach
has been applied to the pseudogap regime of the $2d$ Hubbard model, as observed in DCA. 
It was demonstrated that the development of the pseudogap can be related, in a
complementary description, to the
formation of strong RVB or of antiferromagnetic correlations, 
but {\sl not} to superconductivity or charge fluctuations.

In particular, we have performed DCA calculations up to a 32-site
clusters, which has  three inequivalent ${\bf K}$-points on the Fermi surface. It is 
crucial that different ${\bf K}$-orbitals on the cluster have rather 
different couplings to their baths. Therefore we find that as $U$ is 
increased, correlation functions first obtain values appropriate for 
an RVB state for ${\bf K}=(\pi,0)$ and $(0,\pi)$, which have the weakest 
coupling to their baths. At the same time  pseudogaps form for these 
${\bf K}$-vectors. The strong interaction between the $(\pi,0)$ and 
$(0,\pi)$ for larger values of $U$ is crucial for this result. The 
same happens for ${\bf K} =(\pm 3\pi/4,\pm \pi/4)$ and $(\pm \pi/4,3\pi/4)$ 
for a somewhat larger $U$, due to the stronger coupling to
the corresponding baths. Finally this also happens for ${\bf K}= 
(\pm \pi/2,\pm \pi/2)$ for a still larger $U$, where the coupling to 
the baths are the strongest.

From a purely algorithmic viewpoint, the systematic derivations presented in this
paper demonstrate how it is possible to gain a considerable reduction
of the numerical effort for the fluctuation diagnostic calculations
(see Eq.~(\ref{eq:dse_a}) and the related discussion).

In summary, we clarify the relation between the evolution of spectra and
different correlation functions in the most interesting correlated regime of the $2d$
Hubbard model, where pseudogap features are clearly visible. 
In doing this, we have made contact to two earlier approaches in Refs.~\onlinecite{Fluct} and 
\onlinecite{Jaime}, using the conceptual framework of Eqs.~(\ref{eq:1.1a})
and (\ref{eq:1.2}), respectively. From  our analysis, it becomes
clear how the two approaches indeed capture complementary aspects of the same physics.

\section{Acknowledgments}

We thank M. Capone, P. Chalupa, S. Ciuchi, L. Del Re, J. Le Blanc, and D. Springer 
for insightful discussions. 
J.M. acknowledges financial support from (MAT2012-37263-C02-01,\
MAT2015-66128-R) MINECO/FEDER, EU.
AT and TS  acknowledge financial support from the Austrian Science
Fund (FWF) through the projects SFB-ViCoM F41 (AT) and I 2794-N35 (TS).
TS has received funding from the European Research Council under the European Union’s Seventh Framework Programme (FP7/2007-
2013) ERC Grant Agreement nr. 319286 (Q-MAC).
G.S. acknowledges financial support by the DFG (SFB 1170 “ToCoTronics”)

\appendix\ 
\section{Four-level model. Large $U$ and $\Delta U >0$}\label{sec:A1}

To understand the results for $\Delta U >0$ in a more transparent way, we consider the large $U$ limit.
Then two electrons localize on the cluster, and we can obtain a good 
description of $\Sigma$ by considering the isolated cluster. We furthermore
assume a small $T$ so that only the lowest state is occupied. For $\Delta U <0$,
the system forms a Kondo-like state with the bath, and even for a large $U$ 
we cannot simplify the problem by just considering the isolated cluster. 

Generally, we can write the Green's function for the cluster level 1 as,
\begin{equation}\label{eq:9a}
[g_{1c1c}(\nu)]^{-1}=[g^0_{1c1c}(\nu)]^{-1}-\Sigma_{1c1c \sigma}(\nu),
\end{equation}
where
\begin{equation}\label{eq:9b}
g^0_{1c1c}(\nu)={1\over i\nu-\varepsilon_c-{V^2\over i\nu-\varepsilon_b+\mu}+\mu} 
\end{equation}
and
\begin{eqnarray}\label{eq:14}
&&\Sigma_{1c1c}(\nu) 
=-[Zg_{1c1c}(\nu)]^{-1} \sum_{mn}{e^{-\beta E_m} +e^{-\beta E_n}\over i\nu +E_n-E_m } \nonumber \\
&&\times \langle n|[ -U_{xx}c_{1c\uparrow}n_{1c\downarrow} -U_{xy}c_{1c\uparrow}n_{2c\downarrow} 
-Jc_{1c\downarrow}^{\dagger}c_{2c\downarrow}^{\phantom\dagger}c_{2c\uparrow}^{\phantom\dagger}\nonumber \\  
&&+Jc_{2c\uparrow}^{\dagger}c_{2c\uparrow}^{\phantom\dagger}c_{1c\uparrow}^{\phantom\dagger}
+Jc_{2c\downarrow}^{\dagger}c_{2c\uparrow}^{\phantom\dagger}c_{1c\downarrow}^{\phantom\dagger}]
|m\rangle\langle m|c_{1c\uparrow}^{\dagger}|n\rangle 
\end{eqnarray}
We now consider a large $U/V$ and $\Delta U >0$, so that we can consider the isolated cluster. 
For a small $T$ we obtain                
\begin{eqnarray}\label{eq:29b}
&&\Sigma_{1c1c}(\nu)={1\over 2g_{1c1c}(\nu)}[ {U_{xy}\over i\nu-U/2-3\Delta U/2-J/2}  \nonumber \\
&&+ {U_{xx}\over i\nu+U/2+3\Delta U/2+J/2}      \\
&&-J\lbrace {1\over i\nu+U/2+3\Delta U/2+J/2} \nonumber \\ 
&&-{1\over i\nu-U/2-3\Delta U/2-J/2}\rbrace].\nonumber 
\end{eqnarray}
Here a term proportional to $U_{xy}-J$ has been neglected.
The term
\begin{equation}\label{eq:78}
J\sum_{{\bf k}\ne {\bf k}'} c^{\dagger}_{{\bf k}\uparrow}c^{\dagger}_{{\bf k}\downarrow} 
c^{\phantom \dagger}_{{\bf k}'\downarrow}c^{\phantom \dagger}_{{\bf k}' \uparrow}
\end{equation}
in the Hamiltonian couples the two terms in the simplified ground-state [Eq.~(\ref{eq:5.2a})]
particularly efficient. This leads to the large contribution proportional to $J$ in
Eq.~(\ref{eq:29b}), which corresponds to ${\bf Q}=(\pi,\pi)$ and ${\bf K}_1=(0,\pi)$. 

These results can be directly compared with Fig.~\ref{fig:2}a.
For small values of $\nu$ the first two terms approximately cancel. These two terms 
correspond to the processes ${\bf K_1}=(\pi,0)$ and $(0,\pi)$ for ${\bf Q}=(0,0)$,
which are seen to cancel in Fig.~\ref{fig:2}a. The term proportional to $J$ 
corresponds to ${\bf Q}=(\pi,\pi)$ and ${\bf K}_1=(0,\pi)$. This term contains
two contributions, which approximately add up for small $\nu$. The absolute 
value of this term is then approximately twice as large as the previous two 
terms, as is also found in Fig.~\ref{fig:2}a.

Above we have expressed the self-energy in terms of $\chi_{\uparrow\downarrow}$,
We can instead express it in terms of $\chi_{\rm sp}$, as is done in the 
fluctuation diagnostics.
\begin{eqnarray}\label{eq:6.14}
&&\Sigma_{1c1c}(\nu)={1\over 2g(\nu)}[ 
-U_{xx}{1\over  i\nu-U/2-3\Delta U/2-J/2}\nonumber      \\
&&+J{1\over i\nu-U/2-3\Delta U/2-J/2}       \\
&&-J\lbrace {1\over i\nu+U/2+3\Delta U/2+J/2} \nonumber \\ 
&&-{1\over i\nu-U/2-3\Delta U/2-J/2}\rbrace]\nonumber 
\end{eqnarray}
The first term corresponds to ${\bf Q}=(0,0)$ and ${\bf K}={\bf K}'=(\pi,0)$, the second term to
${\bf Q}=(\pi,\pi)$ and ${\bf K}={\bf K}'=(\pi,0)$ and the third term to ${\bf Q}=(\pi,\pi)$,
${\bf K}=(\pi,0)$ and ${\bf K}'=(0,\pi)$.  This is the largest term, showing the importance of 
${\bf Q}=(\pi,\pi)$ in the fluctuation diagnostics. The first two
terms largely cancel each other. \\

\section{Accuracy of large $U$ approximation}\label{sec:5.6}

Throughout our work, we have extensively used the large $U$ approximation Eq.~(\ref{eq:16a}) in
Eqs.~(\ref{eq:19c}, \ref{eq:25}, \ref{eq:19g}-\ref{eq:19i}) to obtain a relation 
between spectra and real space correlation functions for a half-filled system. 
The accuracy of these approximations is therefore crucial. The approximations 
were of two types. First we made an approximation for the eigenenergies in 
Eq.~(\ref{eq:16a}) and in the following equations we also assumed that double 
occupancy can be neglected. 

We can now discuss the accuracy of these approximations 
by comparing general results of the approximations with results in Figs.~\ref{fig:1}, 
\ref{fig:3}, \ref{fig:5} and \ref{fig:8}, which did not use large $U$ approximations. 
The large $U$ approximations predict that $D_{\uparrow\downarrow}\to -1$ and that
$B_{\uparrow\downarrow}[{\bf K},{\bf K}+(\pi,\pi)]$ are independent of ${\bf K}$.
We first discuss this for K-points at the Fermi surface. The large U results for 
$B$ and $D$ then show up in the Figs. \ref{fig:3}, \ref{fig:5} and \ref{fig:8} 
for values of $U$ where the (pseudo)gap has developed. This happens for $U\sim 1.2$ eV 
($N_c=4$), 1.8 eV ($N_c=8$) and 1.1-1.2 eV ($N_c=32$), in all cases for $t=-0.25$ eV 
and $\beta=60$ eV$^{-1}$. For somewhat smaller values of $U$, $B_{\uparrow\downarrow}
[{\bf K},{\bf K}+(\pi,\pi)]$ has an important dependence on ${\bf K}$, also for ${\bf K}$
at the Fermi surface. For the ${\bf K}$-points away from the Fermi surface the large-$U$ 
approximation becomes valid only for substantially larger values of $U$ than above. 
We have performed calculations for an isolated cluster, since for large $U$ the coupling 
to the bath plays a small role.  For $N_c=8$ we find that $B_{\uparrow\downarrow}
[{\bf K},{\bf K}+(\pi,\pi)]$ becomes independent of ${\bf K}$ for all values of 
${\bf K}$ for $U\!\simeq\!10$ and then $B\!\simeq\!-0.7$. We furthermore find 
that $D_{\uparrow\downarrow}\!\simeq\!-1$ for all values of ${\bf K}$ for $U\!\simeq\!10$. At this point 
the RVB state is fully developed.

As $U$ is increased, $B_{\uparrow\downarrow}[{\bf K},{\bf K}+(\pi,\pi)]$
first approaches large negative values, as in the RVB state, for ${\bf K}$
at the antinodal point [$(\pi,0)$]. As $U$ is further increased this
gradually happens for ${\bf K}$-vectors moving towards the nodal point. 
This happens for moderate values of $U$ and it has above been referred 
to as the formation of an RVB-like state. To fully form an RVB state, however,
$U$ has to be increased substantially more.



\begin{thebibliography}{1}
\bibitem{su2note} In the SU(2) symmetric case the correlation functions $B_{\uparrow\downarrow}(\mathbf{k},\mathbf{K}')$ and $C_{\text{sp}}(\mathbf{K},\mathbf{Q})$ are directly related, which can be easily seen in the following way: Considering the definition of $C_{\text{sp}}(\mathbf{K},\mathbf{Q})$ in Eqs.~(\ref{eq:11c}) and (\ref{eq:19f}) as well as the SU(2) symmetry represented by the relation\cite{Rohringer} $A_{\uparrow\downarrow}(\mathbf{K},\mathbf{K}',\mathbf{Q})=A_{\uparrow\uparrow}(\mathbf{K},\mathbf{K}',\mathbf{Q})+A_{\uparrow\downarrow}(\mathbf{K},\mathbf{K}+\mathbf{Q},\mathbf{K}'-\mathbf{K})$ we immediately obtain that $C_{\text{sp}}(\mathbf{K},\mathbf{Q})=B_{\uparrow\downarrow}(\mathbf{K},\mathbf{K}+\mathbf{Q})$. From a physical perspective, $C_{\text{sp}}$ describes {\sl longitudinal} and $B_{\uparrow\downarrow}$ {\sl transverse} spin fluctuations which are of course the same in a SU(2) symmetric situation.
\bibitem{noteSU2vio} It is interesting to note that in the previous
  case ($\Delta U >0$) the mentioned  equivalence between $B$ and $C$ appears to be
   well preserved (cf. the corresponding values in Table~\ref{table:2}), although
  the corresponding four-level model is {\sl not} SU(2)-symmetric. This
  depends, most likely, on the fact that, for $\Delta U >0$, the (singlet) nature of the
  ground state of the model does not change w.r.t. the
  SU(2)-symmetric case ($\Delta U =0$), differently from what happens in the case
  $\Delta U < 0$.
\bibitem{notelevels} Note, however, that by further increasing $U$,
  i.e., for $U \rightarrow \infty$, all levels will be occupied 
equally, and Eq.~(\ref{eq:5.2}) is not a good approximation any more. Then the four-site 
model becomes quite different from the four-level model.
\end{thebibliography}
\end{document}